\documentclass[pdflatex,sn-mathphys-num]{sn-jnl}

\usepackage{graphicx}%
\usepackage{multirow}%
\usepackage{amsmath,amssymb,amsfonts}%
\usepackage{amsthm}%
\usepackage{mathrsfs}%
\usepackage[title]{appendix}%
\usepackage{xcolor}%
\usepackage{textcomp}%
\usepackage{manyfoot}%
\usepackage{booktabs}%
\usepackage{algorithm}%
\usepackage{algorithmicx}%
\usepackage{algpseudocode}%
\usepackage{listings}%
\usepackage{mathtools}

\usepackage{color,xcolor,ucs}
\usepackage{mathtools}   \usepackage{tikz} 
\usepackage{ amssymb }
\usepackage{extarrows} 
\usepackage{pgf,tikz}
\usepackage{float}
\usetikzlibrary{positioning}
\usetikzlibrary{shapes.geometric}
\usetikzlibrary{shapes.misc}
\usetikzlibrary{arrows}
\usepackage{caption}
\usepackage{mathrsfs}
\usetikzlibrary{arrows,shapes,automata,backgrounds,petri,positioning}
\usetikzlibrary{decorations.pathmorphing}
\usetikzlibrary{decorations.shapes}
\usetikzlibrary{decorations.text}
\usetikzlibrary{decorations.fractals}
\usetikzlibrary{decorations.footprints}
\usetikzlibrary{shadows}
\usetikzlibrary{calc}
\usetikzlibrary{spy}
\usepackage{amsmath}
\usepackage{array}
\usepackage{ amssymb }
\usepackage{braket}
\usepackage{qcircuit}
\usepackage{soul}
\usepackage{braket} 
\usepackage{relsize}

\usepackage{amsmath}
\usepackage{ amssymb }
\usepackage{braket}
\usepackage{qcircuit}
\usepackage{soul}
\usepackage{braket}

\theoremstyle{thmstyleone}%
\theoremstyle{thmstyletwo}%

\theoremstyle{thmstylethree}%

\begin{document}

\title[Article Title]{Composable, unconditional security without a Quantum secret key: public broadcast channels and their conceptualizations, adaptive bit transmission rates, fidelity pruning under wiretaps}


\author[1]{\fnm{Pete} \sur{Rigas}}\email{pbr43@cornell.edu}

\affil[1]{\city{Newport Beach}, \postcode{92625}, \state{CA}, \country{United States}}


\abstract{We examine public broadcast, forward conceptual, and backward conceptual, Quantum channels in the context of communication protocols that are independent of secret keys. Given research directions of interest previously identified in arXiv: 1804.01797, besides converse upper bounds on the bit transmission rate obtained by the author in recent work (arXiv: 2507.03035), additional possibilities remain, including: (1) determining whether aspects of QKD dependent protocols can be incorporated into steps of QKD independent protocols; (2) whether there would be any amplification to the Quantum-classical performance gap that Alice and Bob can exploit towards prospective Quantum advantage; (3) formulating the conditions under which secrecy and authentication can be simultaneously achieved. To characterize the conditions for which secrecy can be achieved with high probability, we argue that there not only exists suitable protocols which enable Alice and Bob to map into the authenticated space of bit codewords with high probability, but also that forward conceptual channels, through cascading, can significantly increase Eve's probability of false acceptance. Albeit the fact that secrecy, along with conceputalizations of the public broadcast channel, were initially discussed by Maurer for QKD dependent protocols, determining whether aspects of such protocols can be adapted for unconditional security without the use of a secret key is of great interest to explore. We demonstrate that Eve's error probability, through the cascading procedure, can be analyzed with the Holevo information under an optimal decoder. Furthermore, through post-processing of the outputs of a Completely Positive Trace Preserving (CPTP) map, we also demonstrate how to decrease Holevo sum quantities with data-processing and entropy-continuity bounds. \footnote{\textbf{MSC Class}: 81P02; 81Q02}}




\maketitle

\section{Introduction}

\subsection{Overview}

\noindent Quantum information theory is a broadly encompassing field, with recent efforts devoted towards comparisons between one way, and two way, communication, {\color{blue}[1]}, variational inference, {\color{blue}[6]}, computational complexity of training {\color{blue}[7]}, discriminant analysis, {\color{blue}[10]}, and several closely related topics {\color{blue}[3},{\color{blue}4},{\color{blue}5},{\color{blue}8},{\color{blue}9},{\color{blue}11},{\color{blue}12},{\color{blue}14},{\color{blue}15},{\color{blue}16},{\color{blue}17},{\color{blue}18},{\color{blue}19},{\color{blue}20]}. In a few previous works from the author this year, several characterizations of the advantages, and correspondingly, limitations, of Quantum information processing protocols have been formally developed, particularly when it comes to optimality, {\color{blue}[47]}, error correction, authentication, and false acceptance, probabilities, {\color{blue}[48]}, in addition to multiplayer parallel repetition {\color{blue}[49]}. While these works are extremely informative in determining the limitations of protocols within the field of Quantum information theory, along with the expected speedups in algorithmic runtime, whether polynomial, or less realistically, exponential, generalizing the settings under which noise can impact the performance of such protocols remains to be of great interest to further explore.

To contribute to a field with rapid theoretical and experimental developments, {\color{blue}[21},{\color{blue}22},{\color{blue}23},{\color{blue}24},{\color{blue}25},{\color{blue}26},{\color{blue}27},{\color{blue}28},{\color{blue}29},{\color{blue}31},{\color{blue}34]}, we begin by describing how generalizations on noise models of Quantum channels can be conceptualized. In such models, classical, or Quantum, information can be distributed across a Quantum channel and one would like to characterize how noise can impact information-theoretic aspects of the bits, or qubits, before, and after, an encoding is transmitted. By \textit{encoding}, and \textit{transmission}, we mean the arrangements of bits, or qubits, that one participant decides to distribute, or to \textit{transmit}, over a Quantum channel that another participant can receive, and further manipulate. Before work of the author this year, {\color{blue}[47]}, which sought to provide converse upper bounds on the bit transmission rate, work due to Ostrev, {\color{blue}[39]}, described how composable, unconditional, security can be transmitted over a Quantum channel \textit{without} the use of a secret key. This work was of great interest to initially explore not only because Quantum key distribution could theoretically outperform protocols associated with classical key distribution, but also because Ostrev claimed that the rate, and converse rate, at which bit can be transmitted over a Quantum channel is dependent upon an artifact of the proof technique, and not upon any 'fundamental physical limit' {\color{blue}[39]}.

With several paradoxical aspects of Quantum information in mind, which raise implications for nonlocality, {\color{blue}[2]}, quantifying the behavior of correlations {\color{blue}[32]}, reducing error rates {\color{blue}[33]}, and even virtual screens for discovery of new therapeutics and materials, {\color{blue}[34},{\color{blue}36},{\color{blue}37},{\color{blue}41]}, addressing other aspects of open problems raised by Ostrev in {\color{blue}[39]} are valuable to explore. Specifically, other questions rather than those pertaining to converse upper bounds on the bit transmission rate, which share connections with game-theoretic settings with two, or an arbitrary number, of participants, {\color{blue}[38},{\color{blue}42},{\color{blue}43},{\color{blue}44},{\color{blue}45]}, surround the \textit{secrecy} capacity, and two-way communication between Bob and Alice. In the absence of two-way, or feedback, communication through a public broadcast channel raises several implications for conditional, and unconditional, security as described in {\color{blue}[35]}. Unexpectedly, even when Eve's channel is \textit{superior}, ie has less noise than Alice and Bob's channels, the presence of two-way communication, as conceptualized in {\color{blue}[35]}, over the public broadcast channel can lead to Alice and Bob generating an information-theoretically secure key. Moreover, besides being able to generate such a key in the first place, the bits which Alice and Bob share to construct such a key are \textit{unconditionally} secure, meaning that the is \textit{no} probability of Eve being able to \textit{break} the key constructed from two-way discussion between Alice and Bob. Explicitly, the aforementioned probability takes the form,

\begin{align*}
   \underset{\textit{bits} \longrightarrow + \infty}{\mathrm{lim}} \textbf{P} \big[   \textit{Eve breaks a finitely many bit key constructed by Alice and Bob th-} \\ \textit{rough public discussion}       \big] \equiv 0   \\ \equiv   \underset{n \longrightarrow + \infty}{\mathrm{lim}} \textbf{P} \big[   \textit{Eve breaks an n-bit key constructed by Alice and Bob through public dis-} \\ \textit{cussion}       \big] \equiv 0  \\ \Longleftrightarrow   \textbf{P} \big[           \textit{Eve breaks an n-bit key constructed by Alice and Bob through public dis-} \\ \textit{cussion}    \big] \equiv 0     . 
\end{align*}

\noindent While there are theoretical possibilities that Alice and Bob can share information, as instances of unconditional security, through the generation of Quantum keys, they can also generate bits, independently of a Quantum key, that can be shared between them that Eve cannot tamper. In the former possibility, the \textit{secrecy capacity} was introduced by Maurer, first \textit{without} feedback, and then \textit{with} feedback. Two such \textit{secret} capacities, respectively denoted with $C_s$ and $\bar{C_s}$, reflect upon the impact of \textit{two-way} communication between Alice and Bob. Despite the fact that \textit{one-way} communication was considered in {\color{blue}[39]}, and subsequently in {\color{blue}[48]}, incorporating aspects of \textit{two-way} communication within a Quantum protocol can not only allow Alice and Bob to repeatedly share, potentially, an arbitrary number of Quantum states, but also opportunities to generate Quantum keys under unconditional security assumptions.

When determining whether a Quantum protocol can generate keys, or bit codewords (as generated by protocols which do not generate secret keys as in {\color{blue}[39},{\color{blue}48]}), with conditional, or unconditional, security guarantees, regardless one must consider the \textit{aggregate} number of bits that can be transmitted per independent channel use. Namely, after each consecutive use of a Quantum channel, the otal number of bits that can be transmitted is dependent upon, as discussed by the author in {\color{blue}[48]}, a \textit{pruning procedure}, that is dependent upon the conditional probabilities,

\begin{align*}
    p_{\mathrm{FA}} \big( \mathscr{O} \big( \textbf{X}, \textbf{Y}, \textbf{Z} \big) \big) \equiv   \underset{\mathrm{fa} \in \mathrm{FA}}{\bigcup}  p_{\mathrm{fa}} \big( \mathscr{O} \big( \textbf{X}, \textbf{Y}, \textbf{Z} \big)    \big) \equiv \underset{\mathrm{fa} \in \mathrm{FA}}{\bigcup}  \big\{ \textbf{P} \big[  \big\{        \text{Alice or Bob commit an in-} \\ \text{stance fa of false acceptance}      \big\} \big|   \big\{  \mathscr{O} \big( \textbf{X}, \textbf{Y}, \textbf{Z} \big) \equiv c^{*}  \big\} \big] \big\}  \equiv 0  \text{, } \\   \\  p_{\mathrm{DE}} \big( \mathscr{O} \big( \textbf{X}, \textbf{Y}, \textbf{Z} \big) \big) \equiv   \underset{\mathrm{de} \in \mathrm{DE}}{\bigcup}  p_{\mathrm{de}} \big( \mathscr{O} \big( \textbf{X}, \textbf{Y}, \textbf{Z} \big)    \big) \equiv   \underset{\mathrm{de} \in \mathrm{DE}}{\bigcup}  \big\{     \textbf{P} \big[  \big\{        \text{Alice or Bob commit an in-} \\ \text{stance de of a decoding error}      \big\} \big|   \big\{  \mathscr{O} \big( \textbf{X}, \textbf{Y}, \textbf{Z} \big) \equiv c^{*}  \big\} \big]     \big\} \equiv 0 , 
\end{align*}

\noindent respectively corresponding to false acceptance, and decoding errors, given the overlap function,

\begin{align*}
   \mathscr{O} \big( \textbf{X}, \textbf{Y}, \textbf{Z} \big) \equiv  \underset{ x \in \textbf{X}, y \in \textbf{Y}, z \in  \textbf{Z}}{\bigcup} \mathscr{O} \big( x, y, z \big) \text{, }
\end{align*}

\noindent given the alphabets $\textbf{X}$, $\textbf{Y}$, and $\textbf{Z}$, of Alice, Bob and Eve, respectively. $\mathscr{O}$, as being defuned through a function of each player's letters, satisfies,

\begin{align*}
  \mathscr{O} \big( \textbf{X}, \textbf{Y}, \textbf{Z} \big) \equiv \emptyset   \Longleftrightarrow  \underset{z \in \textbf{Z}}{\bigcap} \bigg\{             \big\{   z \big\}  \cap \bigg\{ \underset{y \in \textbf{Y}}{\bigcap} \big\{   y \big\}  \cap \big\{  \textbf{X} \big\}  \bigg\}  \bigg\}     \equiv \emptyset  \Longleftrightarrow  \underset{z \in \textbf{Z}}{\bigcap} \bigg\{             \big\{   z \big\} \\  \cap \bigg\{ \underset{x \in \textbf{X}}{\bigcap} \big\{   \textbf{Y} \big\}  \cap \big\{  x \big\}  \bigg\}  \bigg\}   \equiv \emptyset  \text{. }
\end{align*}

\noindent By making use of the above properties of $\mathscr{O}$, one possible converse upper bound for the bit transmission rate $r$ was derived by the author in {\color{blue}[48]}, which takes the form,

\begin{align*}
\underset{P_{\textbf{X}}}{\mathrm{sup}}  \big\{ \mathrm{min} \big\{  I \big( \textbf{X} , \textbf{Y} \big)  , \underset{z}{\mathrm{min}}  \big\{ H_Q \big( \textbf{Y} \big| \textbf{Z} = z \big)   -  H_P \big( \textbf{Y} \big| \textbf{X} \big)  \big\}  \big\}    \big\} 
\end{align*}
\[ <  \left\{\!\begin{array}{ll@{}>{{}}l} 
     \mathrm{log} \mathrm{log} \bigg[   \frac{  \mathrm{log} \big| \textbf{Y}^{*} \big|   }{ \big| \textbf{X}^{*} \big|  }          \bigg]   +   \mathrm{log}  \bigg[   \frac{   \mathrm{log}\big|  \textbf{Z}  \big|  }{  \big|  \textbf{Y}^{*}  \big|   } \bigg]  \Longleftrightarrow \big| \textbf{X} \big| > \big| \textbf{Y}^{*} \big| ,  \big| \textbf{Y}^{*} \big| > \big| \textbf{Z} \big|    ,     \\ \mathrm{log} \mathrm{log} \bigg[     \frac{  \mathrm{log} \big| \textbf{X} \big|  }{ \big| \textbf{Y}^{*} \big| }          \bigg]   +   \mathrm{log}  \bigg[  \frac{ \mathrm{log} \big|  \textbf{Y}^{*} \big|   }{  \big| \textbf{X} \big|   } \bigg]   \Longleftrightarrow \big| \textbf{X} \big| <  \big| \textbf{Y}^{*} \big| ,  \big|  \textbf{Y}^{*} \big| < \big| \textbf{Z} \big|        ,  \\ \mathrm{log} \mathrm{log} \bigg[     \frac{  \mathrm{log} \big| \textbf{Y}^{*} \big|  }{ \big| \textbf{X}^{*} \big| }          \bigg]   +   \mathrm{log}  \bigg[  \frac{  \mathrm{log}\big|  \textbf{Y}^{*} \big|   }{ \big| \textbf{X} \big|   } \bigg]   \Longleftrightarrow \big| \textbf{X} \big| >   \big| \textbf{Y}^{*} \big| ,  \big|  \textbf{Y}^{*} \big| < \big| \textbf{Z} \big|        ,     \\ \mathrm{log} \mathrm{log} \bigg[     \frac{  \mathrm{log} \big| \textbf{Y}^{*} \big|  }{\big| \textbf{X}^{*} \big| }          \bigg]   +   \mathrm{log}  \bigg[  \frac{  \mathrm{log}\big|  \textbf{Z}  \big|   }{ \big| \textbf{Y}^{*} \big|   } \bigg]   \Longleftrightarrow \big| \textbf{X} \big| <    \big| \textbf{Y}^{*} \big| ,  \big|  \textbf{Y}^{*} \big| >   \big| \textbf{Z} \big|        .                          
\end{array}\right. \equiv  r  , 
\]

\noindent for, 

\begin{align*}
    \textit{Mutual Information entropy} \equiv I \big( \cdot , \cdot \big) , 
\\ \\ \textit{Conditional Shannon entropy} \equiv H \big( \cdot \big| \cdot \big) , \\ \\ \textit{Alphabet of the first player, Bob} \equiv \textbf{X} , \\  \\ \textit{Pruned alphabet of the first player, Bob} \equiv \textbf{X}^{*} , \\  \\ \textit{Alphabet of the second player, Alice} \equiv \textbf{Y} , \\ \\ \textit{Pruned alphabet of the second player, Alice} \equiv \textbf{Y}^{*} ,\\  \\ \textit{Alphabet of the third player, Eve} \equiv \textbf{Z} . 
\end{align*}

\noindent If one would like to incorporate two-way communication into Quantum protocols associated with $r$, so that bit codewords shared between Alice and Bob are \textit{unconditionally} secure, as a starting point one would think to perform error correction on the total number of qubits transmitted per independent use of a channel, according to the above $r$. Fortunately, the fact that the above possible converse upper bounds for $r$, of the form 'loglog$+$log', in itself requires that Alice or Bob allocate codewords of substantial length for transmission; moreover, being able to successfully perform error correction on even a \textit{fraction} of the qubits required for satisfying the overhead stipulated by $r$ can  provide ample opportunity towards constructing a shared key, or codeword, between them. Furthermore, albeit the fact that Alice and Bob can, theoretically speaking, exchange keys dependent upon many bits repeatedly through the public broadcast channel, Eve cannot determine any of the bits in the key with unlimited computational power.

To further examine the steps in protocols associated with composable, and unconditional, security, we consider the following programme: (1) we describe how game-theoretic objects, along the lines of those introduced first in {\color{blue}[38]}, and further studied later in {\color{blue}[45},{\color{blue}47},{\color{blue}49]}, are related to the transmission of classical bits over Quantum channels; (2) after having motivated the importance of two-player game-theoretic objects, we describe how multiplayer game-theoretic objects can be straightforwardly conceptualized in order to describe how Eve, as a third participant, can intercept the classical bits transmitted between Alice and Bob; (3) we describe how aspects of cryptography, from discussions in {\color{blue}[39]}, can be used to describe how broadcast channels, through public discussion, ie \textit{two-way} discussion between Alice and Bob, can impact the probabilities of false acceptance, authentication, and decoding errors considered by the author in {\color{blue}[48]}; (4) lastly, determining how objects introduced in {\color{blue}[35]}, for the purposes of unconditional security \textit{with} the use of Quantum secret keys, can be related to protocols discussed in {\color{blue}[39},{\color{blue}48]}, which provide unconditionally security guarantees \textit{without} the use of Quantum secret keys. The last item described in the previous sentence, (4), can be thought of as the primary novelty of this work. In light of the fact that one one can consider Quantum communication protocols that provide \textit{unconditional} security guarantees, it remains of interest, along the open problems suggested in {\color{blue}[35},{\color{blue}39]}, as to whether more possibilities of interest could be studied with \textit{two-way} communication. 

To determine which aspects of Quantum communication protocols are useful towards answering such a research question, one must consider how equalities, and upper bounds for the secrecy capacities without, and with, public discussion were initially introduced in {\color{blue}[35]}. In that conference proceeding, Maurer describes the expected advantages and tradeoffs of incorporating discussion through a public broadcast channel, including: (1) aspects of \textit{unconditional} security, specifically from the fact that Eve cannot break the key's security that is generated through public discussion between Alice and Bob; (2) the fact that a \textit{conceptual} broadcast channel can be introduced, insofar as to increase the error probability of Eve. Conceptualizations of the broadcast channel not only increases the probability that Eve accepts a bit of the secret key transmitted by Alice or Bob over the main channel, but also no longer satisfies, $N_A > N_E$ and $N_B > N_E$.

The noise parameters $N_A, N_B, N_E$ respectively denote the presence of noise of Alice, Bob and Eve's channels. In comparison to noise models appearing in {\color{blue}[39},{\color{blue}48]}, one possible generalization is to suppose that Alice, Bob, and Eve, each have separated channels that they can receive messages over, which are encoded by another participant. Straightforwardly, the noise assumtpion in {\color{blue}[48]} between the noise of the channel shared by Alice and Bob, and between Bob and Eve, $N_{A \longleftrightarrow B} > N_{B \longleftrightarrow E}$, can be generalized to two requirements, including $N_A > N_E$ and $N_B > N_E$. Despite initially appearing as a minor change, imposing  parameters corresponding to the noise of \textit{individual} Quantum channel for each participant allows for concepualizations of the broadcast channel, namely through the backwards broadcast channel. Altogether, it would be of great interest to incorporate broadcast channels, and their conceptualizations, for  composable, unconditionally secure, protocols as examined in {\color{blue}[39},{\color{blue}48]}. Although it is straightforward for one to expect that the \textit{conceptual} broadcast channel can be incorporated into Quantum communication protocols which make \textit{no} use of secret keys, determining whether aspects of protocols which make use of secret keys is of interest to study.

More generally, such a research direction of interest is related to whether Eve, with unlimited computational power, can break \textit{unconditional} security through the Quantum key, or bit codewords, transmitted between Alice and Bob. In otherwords, one can consider whether Eve can determine \textit{any} aspects of Quantum information shared between Alice and Bob which constitute \textit{unconditional} security. While all instances of Quantum communication protocols for which there is \textit{unconditional} security are not necessarily obtained through the use of secret keys, determining whether Maurer's conceptualized broadcast channel can be adopted for communications that are \textit{independent} of secret keys would be of particular interest for the following reasons, including: (1) developing 'hybrid' Quantum communication protocols from instances of \textit{unconditional} security, irrespective of whether a secret key is used; (2) determining whether converse results, such as those proposed for the bit transmission rate $r$, {\color{blue}[39]}, and subsequently formalized by the author in {\color{blue}[48]}, could be leveraged for realizing secrecy; (3) quantifying how capacity related interpretations, first through the total number of bits initially transmitted through the main Quantum channel, and afterwards through the total number of bits that can be communicated between Alice and Bob in secrecy, can be consolidated.

Pertaining to (3) mentioned above, in {\color{blue}[48]} previous arguments of the author determined suitable properties of error correcting codes for minimizing the probability of an instance of false acceptance, while maximizing the probability of an instance of a decoding error. However, despite having such characteristics of error correcting codes in place, it continues to remain of interest to determine, along the lines of communication protocols which \textit{are} dependent upon secret keys, whether expressions for the secrecy capacity and public broadcast channels can be incorporated into error correcting codes for communication protocols that are \textit{independent} of secret keys. To this end, in the remaining sections of this paper, we devote our attention towards:

\begin{itemize}
        \item[$\bullet$] \textit{Bridging the gap between dependent, and independent, secret key communication protocols}. Protocols described in the above comments differ primarily in whether it makes use of a secret key or not. Determining whether there are physical differences in making use of the capacity related result, from previous work of the author on $r$, {\color{blue}[48]}, with public broadcast channels and their conceptualizations, {\color{blue}[35]}, would be of interest for near term applications on new devices within the next fear years. While algorithms for generating Quantum key distribution, as alluded to in {\color{blue}[38},{\color{blue}39]}, could be replaced with more efficient two-hasing protocols could make the implementation of such communication protocols much more feasible within the next few years. Irrespective of the fact that security characterizations of Quantum communication protocols studied in this work at first do not relate to arguments for converse results on $r$, {\color{blue}[48]}, the total number of bits transmitted per channel use is important for protocols \textit{with}, and \textit{without}, use of secret keys. 

          \item[$\bullet$] \textit{Superiority of Eve's independent channel}. In Quantum communication protocols examined in {\color{blue}[35]}, and later in {\color{blue}[39},{\color{blue}48]}, Eve's channel is assumed to be superior to that of Alice, and of Bob. Paradoxically, while it may appear that this assumption on the presence of noise over each Quantum channel would prohibit Alice and Bob from realizing Quantum advantage, such advantage can still be realized. Generally speaking, there are different aspects of a communication protocol which can be exploited for describing aspects of error correcting codes, or other objects of interest in Quantum information theory, which are resilient to noise. 

            \item[$\bullet$] \textit{Algorithmic hybridization}. The fact that the total number of classical bits that can be transmitted over a Quantum channel limits the number of \textit{error corrected} qubits in protocols which \textit{do}, and \textit{do not}, use secret keys establishes further possibilities for two-way communication. Across the public broadcast channel, if Alice (resp. Bob) transmits bits for a secret key  codewords to Bob (resp. Alice), then Bob (resp. Alice) can transmit another collection of bits at capacity $C_s$, which was shown to take the form, (\textbf{Theorem} \textit{1}, {\color{blue}[35]})

\[ C_s \equiv  \left\{\!\begin{array}{ll@{}>{{}}l} 
        h \big( \delta \big) - h \big( \epsilon \big), \text{ } if \text{ } \delta > \epsilon , \\ 0, \text{ } otherwise ,                  
\end{array}\right. 
\] 

\noindent where $h$ denotes the binary entropy function,

\begin{align*}
    h \big( x \big) = x \mathrm{log} x + \big( 1 - x \big) \mathrm{log} \big( 1 - x \big) , 
\end{align*}

\noindent given some $0 < x < 1$, and,

\begin{align*}
  h \big( 0 \big) = h \big( 1 \big) = 0  . 
\end{align*}

\noindent Given that arguments in {\color{blue}[35]} for the \textit{secrecy} capacity, along with the same capacity with discussion over the public broadcast channel, at first sight it appears nontrivial to determine which possibilities two-way communication that the public broadcast channel would imply. One can begin to realize the differences between capacities, related to the \textit{secrecy} capacity, by recalling that the bit transmission rate, besides the specific expression previously obtained by the author,

\[    r \equiv  \left\{\!\begin{array}{ll@{}>{{}}l} 
     \mathrm{log} \mathrm{log} \bigg[   \frac{  \mathrm{log} \big| \textbf{Y}^{*} \big|   }{ \big| \textbf{X}^{*} \big|  }          \bigg]   +   \mathrm{log}  \bigg[   \frac{   \mathrm{log}\big|  \textbf{Z}  \big|  }{  \big|  \textbf{Y}^{*}  \big|   } \bigg]  \Longleftrightarrow \big| \textbf{X} \big| > \big| \textbf{Y}^{*} \big| ,  \big| \textbf{Y}^{*} \big| > \big| \textbf{Z} \big|    ,     \\ \mathrm{log} \mathrm{log} \bigg[     \frac{  \mathrm{log} \big| \textbf{X} \big|  }{ \big| \textbf{Y}^{*} \big| }          \bigg]   +   \mathrm{log}  \bigg[  \frac{ \mathrm{log} \big|  \textbf{Y}^{*} \big|   }{  \big| \textbf{X} \big|   } \bigg]   \Longleftrightarrow \big| \textbf{X} \big| <  \big| \textbf{Y}^{*} \big| ,  \big|  \textbf{Y}^{*} \big| < \big| \textbf{Z} \big|        ,  \\ \mathrm{log} \mathrm{log} \bigg[     \frac{  \mathrm{log} \big| \textbf{Y}^{*} \big|  }{ \big| \textbf{X}^{*} \big| }          \bigg]   +   \mathrm{log}  \bigg[  \frac{  \mathrm{log}\big|  \textbf{Y}^{*} \big|   }{ \big| \textbf{X} \big|   } \bigg]   \Longleftrightarrow \big| \textbf{X} \big| >   \big| \textbf{Y}^{*} \big| ,  \big|  \textbf{Y}^{*} \big| < \big| \textbf{Z} \big|        ,     \\ \mathrm{log} \mathrm{log} \bigg[     \frac{  \mathrm{log} \big| \textbf{Y}^{*} \big|  }{\big| \textbf{X}^{*} \big| }          \bigg]   +   \mathrm{log}  \bigg[  \frac{  \mathrm{log}\big|  \textbf{Z}  \big|   }{ \big| \textbf{Y}^{*} \big|   } \bigg]   \Longleftrightarrow \big| \textbf{X} \big| <    \big| \textbf{Y}^{*} \big| ,  \big|  \textbf{Y}^{*} \big| >   \big| \textbf{Z} \big|        ,                          
\end{array}\right.  
\]

\noindent generally takes the form,

\begin{align*}
\underset{P_{\textbf{X}}}{\mathrm{sup}}  \big\{ \mathrm{min} \big\{  I \big( \textbf{X} , \textbf{Y} \big)  , \underset{z}{\mathrm{min}}  \big\{ H_Q \big( \textbf{Y} \big| \textbf{Z} = z \big)   -  H_P \big( \textbf{Y} \big| \textbf{X} \big)  \big\}  \big\}    \big\} < r , 
\end{align*}

\noindent If one were to obtain a class of suitably defined minimization problems, over probability measures $\textbf{P}_{X}$, permitting \textit{two-way} communication, the accompanying bit transmission rate with discussion, $\bar{r}$, would satisfy,

\begin{align*}
\bar{\underset{P_{\textbf{X}}}{\mathrm{sup}}  \big\{ \mathrm{min} \big\{  I \big( \textbf{X} , \textbf{Y} \big)  , \underset{z}{\mathrm{min}}  \big\{ H_Q \big( \textbf{Y} \big| \textbf{Z} = z \big)   -  H_P \big( \textbf{Y} \big| \textbf{X} \big)  \big\}  \big\}    \big\}} \\ \equiv \underset{P_{\textbf{X}}}{\mathrm{sup}}  \big\{ \mathrm{min} \big\{  \bar{I} \big( \textbf{X} , \textbf{Y} \big)  , \underset{z}{\mathrm{min}}  \big\{ \bar{H_Q} \big( \textbf{Y} \big| \textbf{Z} = z \big)   -  \bar{H_P} \big( \textbf{Y} \big| \textbf{X} \big)  \big\}  \big\}    \big\} < \bar{r} , 
\end{align*}

\noindent where,

\begin{align*}
  \bar{I} \big( \textbf{X} , \textbf{Y} \big)    , \\ \\ \bar{H_Q} \big( \textbf{Y} \big| \textbf{Z} = z \big),  \\ \\  \bar{H_P} \big( \textbf{Y} \big| \textbf{X} \big) , 
\end{align*}

\noindent respectively denote the Mutual Information, and conditional Shannon entropies, with respect to the probability measures $P$ and $Q$, under public discussion.

\bigskip

\noindent Proverbially, while one can straightforwardly define a "counterpart" capacity to the bit transmission rate $r$, with $\bar{r}$, as distributing public discussion over each entry of the minimization, \textit{two-way} communication in communication protocols which do \textit{not} use secret keys provides additional layers of complexity. Such possibilities are related to:

\begin{itemize}
    \item[$\bullet$] \textit{The computation of Mutual Information, and conditional Shannon, entropies under public discussion}. The entropies that one uses to define a capacity related result for $r$, and $\bar{r}$ alike, crucially depend upon well posed constrained optimization problems over probability measures $P_{\textbf{X}}$. Counterintuitively, the way in which one approaches computing the mutual information, and conditional Shannon, entropies, with \textit{or} without discussion through the ordinary and conceptual public broadcast channel are dependent upon \textit{adaptive} bit transmission rates. That is, instead of Alice and Bob making use of a \textit{pruning} procedure, as first identified by the author in {\color{blue}[48]}, Alice and Bob can make use of \textit{variable} bit transmission rates to increase the probability of Eve committing errors.  

      \item[$\bullet$] \textit{Adaptive bit transmission rates over codeword public broadcast, forward conceptual, and backward conceptual, channels}. One can adapt the general setting discussed in {\color{blue}[35]}, and further described in previous remarks, by: specifying how the bit transmission rate over the Main Quantum channel between Alice and Bob, should relate to accompanying transmission rates over the public broadcast, forward conceptual, and backward conceptual, channels; identifying possible manners to increase the error probability of Eve as much as possible over her conceptual public broadcast channel; relating the bits that best to send to Eve; establishing broader comparisons between \textit{pruning} and the error probability of Eve over her conceptual public broadcast channel; more broadly, describing how new possibilities arise from previously examined game-theoretic settings, specifically in the error probabilities of Eve. Even if she initially has an advantage through a lower noise level, and hence lower error acceptance probability over her independent channel, the degree to which this probability can be \textit{increased} is of great interest to formalize. To establish an association between the \textit{decrease} of her error probability over the conceptual public broadcast channel, surprisingly one can make use of communication protocols which \textit{do} make use of secret keys. As a result, converse upper bounds on the bit transmission rate, as first obtained by the author in {\color{blue}[48]}, which were formulated in {\color{blue}[39]}, take on new significance.

        \item[$\bullet$] \textit{Similarities with the secrecy capacity over the backwards conceptual public broadcast channel}. As alluded to in the previous items above, such possibilities arise from incorporating aspects of two-way communication developed in {\color{blue}[35]}. Hence, as expected to some degree, irrespective of whether Alice, or Bob, transmit bit codewords, or bits of a secret key, one can introduce broadcast channels for secret key \textit{independent} protocols. As such, the Quantum communication protocol \textit{independent} of the secret key, if it were to have \textit{unconditional} security guarantees, would still impart prospective Quantum advantage for Alice and Bob inspite of the fact that $N_A > N_E$ and $N_B > N_E$. However, in comparison to forward and backward conceptual channels described in {\color{blue}[35]}, those adapted for secret key \textit{independent} protocols express Eve's error probabilities in terms of,

\begin{align*}
\textit{Alphabet of the first player, Bob} \equiv \textbf{X} , \\  \\ \textit{Pruned alphabet of the first player, Bob} \equiv \textbf{X}^{*} , \\  \\ \textit{Alphabet of the second player, Alice} \equiv \textbf{Y} , \\ \\ \textit{Pruned alphabet of the second player, Alice} \equiv \textbf{Y}^{*} ,\\  \\ \textit{Alphabet of the third player, Eve} \equiv \textbf{Z} . 
\end{align*}

        \noindent which were originally introduced with \textit{pruning}. This procedure was originally motivated to erode Eve's \textit{initial} superiority over Alice and Bob.

\end{itemize}

\end{itemize}

\subsection{This paper's contributions}

\noindent This paper examines novel Quantum communication protocols associated \textit{without} the use of secret keys. Along the lines of previous arguments provided by the author for the bit transmission rate, {\color{blue}[48]}, one can formulate applications that have not been previously examined in the literature, with consequences towards: determining protocols for increasing Eve's error probability; establishing how bit transmission rates should be \textit{adapted} over conceptual forward, and backward, public broadcast channels; discussing how Alice and Bob can increase their prospective Quantum advantage by as much as possible; relating computations for Eve's error probability to those obtained for communication protocols which \textit{do} make use of secret keys {\color{blue}[35]}. More generally, determining the extent to which a Quantum algorithm can outperform a classical algorithm is important for learning which experimental, and possible future commercial, opportunities exist. While scaling up next generation Quantum hardware remains extremely difficult to achieve, theoretically investigating future applications first is much more feasible. Consequentially, the communication protocols described in this work describe settings under which more significant prospective Quantum advantage exists.

We achieve these objectives through extensive manipulation of the CPTP map. Measurement, and post-processing in general, over CPTP maps corresponds to channel degradation which can be leveraged for quantifying how Eve's error probability increases across the cascade. By making use of the MAC, {\color{blue}[52]} and cq-polar codes {\color{blue}[53]}, we introduce Holevo sums, from the Holevo entropy, which take the place of the sum rate that is originally introduced for MACs. However, with Mauerer's public broadcast channel and its associated cascade, we draw the attention of the reader to distillation, in addition to how one-way encodings impact Eve's error probability. To preserve information reconciliation as much as possible, cq-polar codes that have been constructed by Wilde and Guha are used. For such coders, privacy amplification corresponds to polar secure hashing, which in turn can be used to characterize advantage distillation, in which Alice and Bob are able to select polarized bit-channels that are good for them but bad for Eve.

\bigskip

\noindent We make use of the following properties of the Holevo information:

\begin{itemize}
    \item[$\bullet$] \textit{Holevo monotonicity through the data processing inequality},

    \item[$\bullet$] \textit{Continuity of the von Neumann entropy},

    \item[$\bullet$] \textit{Wilde Guha cq-Polarization theorems},

    \item[$\bullet$] \textit{Tal-Sharov-Vardy alphabet reduction},

    \item[$\bullet$] \textit{Privacy amplification through the leakage of Holevo information},

     \item[$\bullet$] \textit{Error analysis for the polar codes},

     \item[$\bullet$] \textit{Wiretaps and their achievability through random coding and privacy amplification}.

\end{itemize}

\subsection{Paper organization}

\noindent After having provided an Overview and description of this paper's contributions in \textit{1.1} and \textit{1.2}, respectively, in the next section we introduce objects associated with two player games, multiplayer games, cryptography, and the public broadcast channel. With such objects, we provide a statement of the main results, relating to: (1) the difference in bit transmission rates over the public broadcast, Eve's forward conceptual channel, and Alice and Bob's backward conceptual channel; (2) obtaining expressions for the Quantum-classical gap between Eve's error probability, through \textit{cascading}, over the forward conceptual channel; (3) expressing Eve's error probability as a function of letters from Alice and Bob's alphabets. We denote the three bit transmission rates described in (1) with,

\begin{align*}
   r \bigg|_{\textit{Public broadcast channel}} \equiv r_1  , \\ \\   r \bigg|_{\textit{Eve's conceptual channel}} \equiv r_2 , \end{align*}
   
   \begin{align*} r \bigg|_{\textit{Alice and Bob's backward conceptual channel}} \equiv r_3 . 
\end{align*}

\noindent With such bit transmission rates, the expected gap, due to cascading, of Eve's error probabilities,

\begin{align*}
  \epsilon_{\textit{Main channel}}  , \epsilon_{\textit{Forward conceptual channel}}  
\end{align*}

\noindent over Eve's main channel, and forward conceptual channel, respectively, imply,

\begin{align*}
    \epsilon_{\textit{Main channel}}  - \epsilon_{\textit{Forward conceptual channel}}  \approx C_{\mathrm{Eve}} ,  \tag{Eve's false acceptance probability gap}
\end{align*}

\noindent for some strictly positive constant, $C_{\mathrm{Eve}}$. In particular, determining how the largest order of approximation between Eve's error probabilities depends upon:

\begin{itemize}
\item[$\bullet$] The number of bits transmitted, $n$, in codewords initially sent by Alice or Bob,

\item[$\bullet$] The encoding employed by Alice or Bob as bits are prepared for each codeword,

\item[$\bullet$] Transmission mechanisms over the Alice's, and Bob's, independent channels, 

\item[$\bullet$] The bit transmission rate, $r$, associated with Alice and Bob's attempts to prevent Eve from tampering with codewords,

\item[$\bullet$] Mappings between the main channel and authenticated channel, as previously characterized by the author in {\color{blue}[48]},

\item[$\bullet$] Hamming ball lower bounds in which the $l$-1 norm of the radius of each such ball is inversely proportional with respect to channel noise,

\item[$\bullet$] Logarithmic factors of,

\begin{align*}
\textbf{X} ,    \textbf{X}^{*} ,    \textbf{Y} ,   \textbf{Y}^{*} ,     \textbf{Z} . 
\end{align*}

\noindent through the appearance of,

{\small \[ \left\{\!\begin{array}{ll@{}>{{}}l}     \mathrm{log} \mathrm{log} \bigg[   \frac{  \mathrm{log} \big| \textbf{Y}^{*} \big|   }{ \big| \textbf{X}^{*} \big|  }          \bigg] ,   \\    \mathrm{log}  \bigg[   \frac{   \mathrm{log}\big|  \textbf{Z}  \big|  }{  \big|  \textbf{Y}^{*}  \big|   } \bigg]    , \\  \\ \mathrm{log} \mathrm{log} \bigg[     \frac{  \mathrm{log} \big| \textbf{X} \big|  }{ \big| \textbf{Y}^{*} \big| }          \bigg]  ,  \\     \mathrm{log}  \bigg[  \frac{ \mathrm{log} \big|  \textbf{Y}^{*} \big|   }{  \big| \textbf{X} \big|   } \bigg]  , \\ \\   \mathrm{log} \mathrm{log} \bigg[     \frac{  \mathrm{log} \big| \textbf{Y}^{*} \big|  }{ \big| \textbf{X}^{*} \big| }          \bigg]  , \\    \mathrm{log}  \bigg[  \frac{  \mathrm{log}\big|  \textbf{Y}^{*} \big|   }{ \big| \textbf{X} \big|   } \bigg]   ,
\\  \mathrm{log} \mathrm{log} \bigg[     \frac{  \mathrm{log} \big| \textbf{Y}^{*} \big|  }{\big| \textbf{X}^{*} \big| }          \bigg]   , \\    \mathrm{log}  \bigg[  \frac{  \mathrm{log}\big|  \textbf{Z}  \big|   }{ \big| \textbf{Y}^{*} \big|   } \bigg]  .             
\end{array}\right.
\]  }



\item[$\bullet$] Stability of encoding, decoding, transmission, and authentication, protocols for infinitely many bits, as previously characterized by the author in {\color{blue}[48]}.

\end{itemize}

\noindent We characterize the implications of bit transmission rates, particularly through the previously defined \textit{adaptive} scheme. Following an overview of the Quantum information-theoretic objects in \textit{2.1}, \textit{2.2}, \textit{2.3}, and \textit{2.4}, we state the main results of this work in \textit{2.5}. Arguments for each main result are provided in \textit{3}, with a conclusion overview of the work provided in \textit{4}.

\section{Quantum information-theoretic objects}

\subsection{Two player game-theoretic objects}

\noindent We define several quantities for an overview of strategies for the infinite $\mathrm{CHSH(n)}$ family of XOR games. First, from the Frobenius norm,

\begin{align*}
  \big|\big| A \big|\big|_F \equiv \sqrt{\overset{m}{\underset{i=1}{\sum}} \overset{n}{\underset{j=1}{\sum}} \big| a_{ij} \big|^2 } = \sqrt{\mathrm{Tr} \big[ A^{\dagger} A \big] }  \text{, } 
\end{align*}

\noindent of an $m \times n$ matrix $A$ with entries $a_{ij}$, there exists a \textit{linear bijection} $\mathcal{L}$ between the tensor product space, $\textbf{C}^{d_A} \otimes \textbf{C}^{d_B}$, and the space of $d_A \times d_B$ matrices with complex entries, $\mathrm{Mat}_{d_A , d_B} \big( \textbf{C} \big)$, satisfying (\textbf{Lemma} \textit{1}, {\color{blue}[38]}),

\begin{itemize}
\item[$\bullet$] \underline{\textit{Image of the tensor product of two quantum states under} $\mathcal{L}$}: $\forall \ket{u} \in \textbf{C}^{d_A}, \ket{w} \in \textbf{C}^{d_B}, \exists \ket{u^{*}} \in \textbf{C}^{d_B} : \mathcal{L} \big( \ket{u} \otimes \ket{w} \big) = \ket{u} \bra{u^{*}}  \text{, }$ 
\item[$\bullet$] \underline{\textit{Product of a matrix with the image of a quantum state under} $\mathcal{L}$}: $\forall \ket{u} \in \textbf{C}^{d_A}, \exists A \in \mathrm{Mat}_{d_A} \big( \textbf{C} \big) : A \mathcal{L} \big( \ket{u} \big) = \mathcal{L} \big( A \otimes I \ket{u} \big)\text{, }$
\item[$\bullet$] \underline{\textit{Product of the image of a quantum state under $\mathcal{L}$ with the transpose of a matrix}}:  $\forall \ket{w} \in \textbf{C}^{d_B}, \exists B \in \mathrm{Mat}_{d_B} \big( \textbf{C} \big) : \mathcal{L} \big( \ket{w} \big) B^T = \mathcal{L} \big( I \otimes B \ket{w} \big)  \text{, }$
\item[$\bullet$] \underline{\textit{Frobenius norm equality}}: $\forall \ket{w} \in \textbf{C}^{d_B} : \big|\big| \mathcal{L} \big(   \ket{w}     \big) 
 \big|\big|_F = \ket{w}  \text{. } $
\end{itemize}

\noindent where the basis of $\textbf{C}^{d_A} \otimes \textbf{C}^{d_B}$ is of the form $\ket{i} \otimes \ket{j}$,  and the basis for $\mathrm{Mat}_{d_A, d_B} \big( \textbf{C} \big)$ is of the form $\ket{i}\bra{j}$, for $1 \leq i \leq d_A$ and $1 \leq j \leq d_B$. From the four properties above of $\mathcal{L}$, for two finite sets $S$ and $T$, also define the map $V : S \times T \longrightarrow \big\{ - 1 , 1 \big\}$. From a product probability distribution $\pi$ over $S \times T$, the game proceeds with the Referee examining the responses of Alice and Bob depending upon the entangled state that they share, in which, after sampling a pair $\big( S , T \big) \sim \pi$, and sending one question $s$ to Alice and another question $t$ to Bob,

\begin{align*}
    V \big(  s , t \big)   ab  \equiv 1 \Longleftrightarrow  \text{ Alice and Bob win,}    \\    V \big(  s , t \big)   ab \equiv -1 \Longleftrightarrow  \text{ Alice and Bob lose,}      
\end{align*}

\noindent in which, depending upon whether $V \big( s ,t \big) \equiv 1$, or $V \big( s , t \big) \equiv -1$, Alice and Bob must either give the same answers, and opposing answers, to win, respectively. The quantities $a$ denote the answer which Alice provides to the Referee after receiving question $s$, while $b$ denotes the answer which Bob provides to the Refree after receiving question $t$. 

Equipped with $V$ and $\pi$, there exists a \textit{game matrix} $G$, so that $G_{st} = V \big( s, t \big) \pi \big( s, t \big)$. Subject to the normalization that the sum over all rows and columns equal $1$, ie $\sum_{st} G_{st} \equiv 1$, a \textit{quantum strategy} for the XOR game is denoted with $\mathcal{S}$, with corresponding state $\ket{\psi} \in \textbf{C}^{d_A} \otimes \textbf{C}^{d_B}$. For an XOR game $G$ and strategy $\mathcal{S}$, define,

\begin{align*}
      \beta \big( G , \mathcal{S}  \big)     \equiv \underset{s \in S}{\sum} \underset{t \in T}{\sum} G_{st} \bra{\psi} A_s \otimes B_t \ket{\psi}  \text{, } 
\end{align*}

\noindent as the \textit{success bias}, where the summation runs over all rows and columns $s$ and $t$ of $G$, with the observables in the tensor product taking the form,

\begin{align*}
 A_S \equiv \underset{s \in S}{\bigcup} A_s \equiv \big\{   s  \in S :  A_s \in \big\{ - 1 , + 1 \big\}    \big\}     \text{, } \\    B_T \equiv \underset{t \in T}{\bigcup} B_t \equiv \big\{ t \in T  :  B_t \in \big\{ -1 , + 1 \big\}  \big\}    \text{. }
\end{align*}

\noindent The quantity above is related to the probability of winning the XOR game given $\mathcal{S}$, denoted as $\omega \big( G , \mathcal{S} \big)$, as,

\begin{align*}
       \beta \big( G , \mathcal{S}  \big)  = 2   \omega \big( G , \mathcal{S} \big) - 1      \text{. } 
\end{align*}

\noindent As a supremum over all possible $\mathcal{S}$ for $G$, define,

\begin{align*}
 \beta \big( G \big) \equiv  \underset{\mathcal{S}}{\mathrm{sup}} \text{ } \beta \big( G , \mathcal{S} \big) \text{, } 
\end{align*}

\noindent corresponding to the optimal quantum strategy. From the optimal strategy $\beta \big( G \big)$, the notion of approximately optimal strategies can be introduced, in which for some strictly positive $\epsilon$,

\begin{align*}
\big( 1 - \epsilon \big) \beta \big( G \big)   \leq   \beta \big( G , \mathcal{S} \big)        \leq  \beta \big( G \big) \text{. }
\end{align*}

\subsection{Multiplayer game-theoretic objects}

\noindent 
\begin{itemize}
    \item[$\bullet$] \underline{\textit{Product norm of player responses}}: Under the identifications,

    \begin{align*}
   \ket{\underset{\text{Odd number of players}}{\prod} i_j } \longleftrightarrow  \underset{\text{Odd number of players}}{\prod}  \ket{ i_j }   \text{, } \\ \bra{\underset{\text{Even number of players}}{\prod} i_j } \longleftrightarrow \underset{\text{Even number of players}}{\prod} \bra{ i_j }   \text{, }
     \end{align*}

    \noindent The outer product of responses from a group of $N$ players can be expressed as,

\begin{align*}
 \bigg[ \ket{\underset{\text{Odd number of players}}{\prod} i_j } \bigg]  \bigg[ \bra{\underset{\text{Even number of players}}{\prod} i_j } \bigg]    \equiv   \bigg[ \ket{\underset{\text{Odd } j, 1 \leq j \leq N}{\prod} i_j } \bigg]  \\ \times \bigg[ \bra{\underset{\text{Even } j, 1 \leq j \leq N}{\prod} i_j } \bigg]   \\ \equiv  \bigg[  \ket{i_N i_{N-2} \times \cdots \times i_1 } \bigg]  \bigg[  \bra{i_2 i_4 \times \cdots \times i_{N-1}} \bigg]    \equiv  \bigg[   \ket{i_N} \ket{i_{N-2}} \times \cdots \times \ket{i_1} \bigg]  \\ \times  \bigg[  \bra{i_2} \bra{i_4} \times \cdots  \times \bra{i_{N-1}} \bigg]  \\  \equiv \ket{i_N}  \bigg[ \cdots \times  \bigg[ \cdots \times \bigg[  \ket{i_3}  \bigg[ \ket{i_1} \bra{i_2} \bigg]   \bra{i_4} \bigg]   \times \cdots  \bigg]  \times  \cdots \bigg]   \bra{i_{N-1}}  \\  \equiv      \ket{\text{Player } N \text{ responds to question } i_N \text{ given } (N-1) \text{ previous responses}}  \bigg[ \cdots  \times \bigg[ \cdots \\  \times \bigg[ \ket{\text{Player } 1 \text{ responds to the first question}}  \bra{\text{Player } 2 \text{ responds to the second question}} \bigg]   \\ \cdots \times \bigg]  \cdots \bigg]      \bra{\text{Player } (N-1) \text{ responds to question } i_N \text{ given } (N-2) \text{ previous responses}}  \text{. }
\end{align*}

\noindent For the $3$ $\mathrm{XOR}$, and $4$ $\mathrm{XOR}$, games, the outer product for $N$ players take the form,

\begin{align*}
\ket{i_3}  \big[ \ket{i_1} \bra{i_2} \big]    \text{, } \\ \ket{i_3}  \big[ \ket{i_1} \bra{i_2} \big]  \bra{i_4} \text{, }
\end{align*}

\noindent respectively.

    \bigskip

     \item[$\bullet$] \underline{\textit{Tensor observables for players of the game}}. To define the multiplayer bias, which will be used to characterize exact, and approximate, optimality up to some parameter $\epsilon$ taken to be sufficiently small, define,

     \begin{align*}
         \bigotimes \text{Player tensor observables} \equiv \bigg[ \text{Alice's observables} \bigg]  \bigotimes \bigg[ \text{Bob's observables} \bigg]  \\ \bigotimes  \bigg[ \text{Cleo's observables} \bigg]    \text{, }
     \end{align*}

     \noindent corresponding to the Hilbert space spanned by the possible set of responses for three players Alice, Bob and Cleo.

\bigskip

      \item[$\bullet$] \underline{\textit{Intertwining operation}}. For tensor products of player observables, error bounds for the two-player $\mathrm{XOR}$ game consist of \textit{interchanging} the order in which the observables that each player forms appear in tensor products, such as the one provided over all player observables above. In error bounds that will follow, denote the intertwining operation, $\widetilde{\cdot}$, where,

\begin{align*}
 \widetilde{\cdot} : M \otimes N \longrightarrow N \otimes M   \text{, }
\end{align*}

\noindent as the permutation operator applied to tensor products $M \otimes N$, where $M$ is a vector with entries from responses of the first player after the referee draws the first question from soem probability distribution of all questions. The above operation is applied under many circumstances, not only for games with more participants but also for games obtained under strong parallel repetition.

\bigskip

         \item[$\bullet$] \underline{\textit{Strong parallel repetition}}. The operation of performing strong parallel repetition, within the exactly, and approximately, optimal framework, is of great interest to further explore and formalize. Under the repetition operation, tensor observables gathered by each player and concatenated together under, potentially, an arbitrary number of games as the referee continues drawing questions from the probability distribution. For any number of players, the strong parallel repetition operation can be straightforwardly extended from two-player settings. In such settings, the action of the strong parallel repetition operation is defined with,

         \begin{align*}
          \text{Strong parallel repetition of Alice's responses to Questions } i \text{ and } j \equiv A_i \wedge A_j   \text{,}
         \end{align*}

         \noindent with the same action being defined for observables gathered by any of the other players.

\bigskip

         \item[$\bullet$] \underline{\textit{$\epsilon$-deviations from optimality}}. Given the existence of a sufficiently small parameter, besides differences in the formulation of error bounds, the bias, and optimal value, satisfying $\epsilon$-approximate optimality, reads,

\begin{align*}
  \big( 1 - \epsilon \big) \beta \big( G \big)  \leq  \underset{\text{Questions}}{\sum} \bra{\text{Optimal Strategy}} \bigg[    \underset{\# \text{ Players }}{\bigotimes}   \text{Tensors of player observables}   \bigg]  \\ \times    \ket{\text{Optimal Strategy}}  \leq      \beta \big( G \big)   \text{, }
\end{align*}

\noindent for a game $G$.
         
\end{itemize}

\subsection{Cryptography theoretic objects}

Adopting the notation originally provided in {\color{blue}[39]}, denote:

\begin{align*}
      \mathcal{R} \neq \mathcal{S} \equiv \text{Two resources which take inputs from Alice, Bob and Eve}    \text{, } \\ \\               \mathcal{R} \big| \big| \mathcal{S} \equiv \text{A resource of } \mathcal{R}  \text{ and } \mathcal{S}  \text{ simultaneously}           \text{, } \\ \\  d \big( \cdot , \cdot \big) \equiv \text{A metric between two resources}     \text{, } \end{align*}

      \begin{align*} \mathcal{N}^n_{p,q} \equiv \text{Alice and Bob's resource for sharing } n\text{-bit authenticated messages over}   \\ \text{the Quantum channel with Bernoulli}-p\text{ random variables}    \text{, }  \end{align*}

      \begin{align*} \mathcal{A}^{rn} \equiv    \text{Alice and Bob's resource for sharing } n\text{-bit authenticated message over the} \\ \text{Quantum channel}       \text{. }
\end{align*}

\noindent Each of the two resources, $\mathcal{R}$ and $\mathcal{S}$, introduced above denote interfaces over the Quantum channel that Alice, Bob, or Eve, can transfer bits into, from which responses can be generated. The metric between resources of the channel, $d \big( \cdot, \cdot \big)$, satisfies, {\color{blue}[39]},

\begin{itemize}
\item[$\bullet$] \textit{Identity}: The distance, with respect to the metric, of a resource with itself, vanishes,

\begin{align*}
  d \big( \mathcal{R}, \mathcal{R} \big) \equiv 0  \text{. }
\end{align*}
\item[$\bullet$] \textit{Symmetry}: The distance, with respect to the metric, between $\mathcal{R}$ and $\mathcal{S}$, and between $\mathcal{S}$ and $\mathcal{R}$, are equal,

\begin{align*}
  d \big( \mathcal{R}, \mathcal{S} \big) \equiv   d \big( \mathcal{S}, \mathcal{R} \big)  \text{. }
\end{align*}

\item[$\bullet$] \textit{The triangle inequality}. Denote three resources over a Quantum channel with $\mathcal{R}$, $\mathcal{S}$, and $\mathcal{T}$. One has that,

\begin{align*}
   d \big( \mathcal{R}, \mathcal{S} \big)    + d \big( \mathcal{S}, \mathcal{T} \big)         \geq  d \big( \mathcal{R}, \mathcal{T} \big)   \text{. }
\end{align*}
\end{itemize}

\noindent Under the assumption that the noise over the Quantum channel between Alice and Bob is strictly less than that over the Quantum channel between Bob and Eve, a version of the stochastic domination, in \textbf{Theorem} \textit{2}, which is related to the converse result, in \textbf{Theorem} \textit{1}, can be anticipated more easily. That is, given the two protocols,

\begin{align*}
  D^n  \overset{n \longrightarrow + \infty}{\longrightarrow} D^{+\infty} \equiv \text{Alice's protocol on codewords with infinitely many bits}  \text{, } \\   E^n   \overset{n \longrightarrow + \infty}{\longrightarrow} E^{+\infty} \equiv \text{Bob's protocol on codewords with infinitely many bits}  \text{, }
\end{align*}

\noindent which are implemented by Alice, and Bob, respectively, for communication over the Quantum channel, denote the resource,

\begin{align*}
  E^n D^n \mathcal{N}^n_{p,q}  \text{, }
\end{align*}

\noindent as that which is obtained by applying Bob's, and Alice's, protocols to the bits that are transmitted over the Quantum channel, in addition to the resource,

\begin{align*}
   \sigma_E \mathcal{A}^{rn} \text{, }
\end{align*}

\noindent obtained by an application of Eve's simulator, $\sigma_E$, to the set of \textit{authenticated} $rn$ bit messages transmitted over the Quantum channel. The distance between resources,

\begin{align*}
    d \big( E^n D^n \mathcal{N}^n_{p,q}, \sigma_E \mathcal{A}^{rn} \big)       \text{, }
\end{align*}

\noindent is identically,

\begin{align*}
  \mathrm{max} \big( p_{\mathrm{de}} , p_{\mathrm{fa}} \big)  \text{, }
\end{align*}

\noindent for,

\begin{align*}
 \textit{Decoding error probability} \equiv p_{\mathrm{de}}    , \\ \\ \textit{False acceptance probability} \equiv  p_{\mathrm{fa}} . 
\end{align*}

\subsection{Capacities rather than those pertaining to converse bit transmission rates: the secrecy capacity over the public broadcast channel}

\noindent We provide a description of the key objects introduced in {\color{blue}[35]} - which have been the subject of discussion in \textit{1.1} and \textit{1.2}.

\bigskip

\noindent \textbf{Theorem} \textit{2}, {\color{blue}[35]} (\textit{upper bound on public discussion of the secrecy capacity}, $\bar{C_s}$, \textit{in terms of the mutual information entropy}, $I$). The secrecy capacity with public discussion of a discrete, memoryless broadcast channel specified by the conditional joint probability measure $\textbf{P}_{YZ|X}$ is upper bounded by,

\begin{align*}
  \bar{C_s} \leq \underset{\textbf{P}_X}{\mathrm{sup}} \big[  I \big( X ; Y | Z \big) \big] \equiv \underset{\textbf{P}_X}{\mathrm{sup}} \big[   H \big( X | Z \big) - H \big( X | YZ \big)     \big]            . 
\end{align*}

\noindent For the next result, denote $\epsilon$ and $\delta$ as two strictly positive parameters. Given previous discussions of the conceptualized broadcast channel, and the backwards broadcast channel, it is natural to consider quantity,

\begin{align*}
    \epsilon + \delta - 2 \epsilon \delta , 
 \end{align*}

\noindent which corresponds to the \textit{cascading} of error probabilities over the conceptual broadcast channel. The quantity defined above, in tandem with the conceptual broadcast channel that Eve uses to receive bits of the secret key with an error probability that is strictly larger than,

\begin{align*}
 \textit{Eve's error probability} \equiv \epsilon_E   ,
\end{align*}

\noindent over her channel, can be used to formulate another expression for $\bar{C_s}$, provided below.

\bigskip

\noindent \textbf{Theorem} \textit{3}, {\color{blue}[35]} (\textit{equality on public discussion of the secrecy capacity, } $\bar{C_s}$, \textit{in terms of the binary entropy function}, $h$, \textit{of the cascaded error probability over the conceptual broadcast channel}). The secrecy capacity with public discussion of the broadcast channel, where the main channel is a binary symmetry channel with error probability $\epsilon \leq \frac{1}{2}$ and Eve's main channel is an independent binary symmetry channel with error probability $\delta \leq \frac{1}{2}$, is given by,

\begin{align*}
   \bar{C_s} = \underset{\textbf{P}_X}{\mathrm{sup}} \big[  I \big( X ; Y | Z \big)    \big] \equiv   h \big( \epsilon + \delta - 2 \epsilon \delta \big) - h \big( \epsilon \big)              .
\end{align*}

\noindent \textbf{Theorem} \textit{4}, {\color{blue}[35]} (\textit{lower bound on public discussion of the secrecy capacity}, $\bar{C_s}$, \textit{in terms of the binary entropy function,} $h$, \textit{of the cascaded error probabilities over the conceptual broadcast channel between Alice and Eve, between Bob and Eve, and between Alice and Bob, respectively}). Denote,

\begin{align*}
   \textit{Alice's error probability}  \equiv  \epsilon_A  , \\ \\ \textit{Bob's error probability}  \equiv  \epsilon_B  , \\ \\ \textit{Eve's error probability}  \equiv  \epsilon_E  , 
\end{align*}

\noindent from which one has that the secrecy capacity with public discussion of the broadcast channel satisfies,

\begin{align*}
 \bar{C_s} \geq \mathrm{sup} \big[ h \big( \epsilon_A + \epsilon_E - 2 \epsilon_A \epsilon_E \big) , h \big( \epsilon_B + \epsilon_E - 2 \epsilon_B \epsilon_E \big)       \big] - h \big( \epsilon_A + \epsilon_B - 2 \epsilon_A \epsilon_B \big)    .
\end{align*}

\noindent Clearly, given that,

\begin{align*}
 h \big( \epsilon \big) < h \big( \epsilon + \delta - 2 \epsilon \delta \big)    , \\ \\     h \big( \delta \big) < h \big( \epsilon + \delta - 2 \epsilon \delta \big)      , 
\end{align*}

\noindent over the conceptual backward channel the secrecy capacity that Alice and Bob can use is dependent upon \textit{smaller} error probabilities that are used as input parameters to the binary entropy function. In contrast to the previous expressions provided for $\bar{C_s}$ in $\textbf{Theorem}$ \textit{2}, \textit{3}, and \textit{4}, we conclude the summary of the main results from {\color{blue}[35]} with $\textbf{Theorem}$ \textit{1}, which states:

\bigskip

\noindent \textbf{Theorem} \textit{1}, {\color{blue}[35]} (\textit{the secrecy capacity independently of public discussion that Alice and Bob can use over the backward conceptual public broadcast channel}). In the absence of public discussion over the broadcast channel, one has that,

\[ C_s \equiv  \left\{\!\begin{array}{ll@{}>{{}}l} 
        h \big( \delta \big) - h \big( \epsilon \big), \text{ } if \text{ } \delta > \epsilon , \\ 0, \text{ } otherwise .                  
\end{array}\right. 
\]

\noindent With several expressions provided for $C_s$, and $\bar{C_s}$, we provide a statement of the main results.

\subsection{Statement of main results}

\noindent The main results include the following assertions:

\begin{itemize}
    \item[$\bullet$] There exists suitable partitions of the letters from Alice and Bob's alphabets which \textit{will} increase Eve's error probability over the forward conceptual channel, 

        \item[$\bullet$] Alice and Bob can maintain superiority, in terms of their respective error probabilities, in comparison to the same probability over Eve's forward conceptual channel, 

            \item[$\bullet$] The expected gap between Eve's error probability over her independent channel, albeit initially being superior to that over Alice and Bob's independent channels, is \textit{amplified} due to the \textit{cascading} of Eve's error probabilities for QKD \textit{dependent} protocols. 
            
            \item[$\bullet$] Similarities with results of previous work of the author provided in {\color{blue}[48]}, which pertain to:
                \begin{itemize}
                    \item[$\bullet$] \textbf{Theorem} \textit{1}, converse result on the bit transmission rate, 
                    \item[$\bullet$] \textbf{Theorem} \textit{2}, stochastic domination between error correction, and false acceptance, probabilities over the forwards and backwards conceptual public broaccast channels, 
                    \item[$\bullet$] \textbf{Theorem} \textit{3}, protocols for mapping into the authenticated space with high probability.
                \end{itemize}
            
            \end{itemize}

\noindent In the following first main result below, introduce the alphabets,

\begin{align*}
  \mathcal{X} \equiv \textit{Alice's alphabet over Eve's forward conceptual channel}  , \\ \\  \mathcal{Y} \equiv \textit{Bob's alphabet over Eve's forward conceptual channel} , \\ \\ \mathscr{X} \equiv \textit{Alice's alphabet over her shared backward conceptual channel with} \\ \textit{ Bob} ,   \\ \\ \mathscr{Y} \equiv \textit{Bob's alphabet over his shared backward conceptual channel with} \\ \textit{Alice} , \\ \\ \mathcal{X}^{*} \equiv \textit{Alice's pruned alphabet over Eve's forward conceptual channel} , \\ \\ \mathcal{Y}^{*} \equiv \textit{Bob's pruned alphabet over Eve's forward conceptual channel} ,  \\ \\ \mathscr{X}^{*} \equiv \textit{Alice's pruned alphabet over her backward conceptual channel with} \\ \textit{Bob} , \\ \\ \mathscr{Y}^{*} \equiv \textit{Bob's pruned alphabet over his backward conceptual channel with} \\ \textit{Alice} ,
\end{align*}

\noindent which satisfy,

\begin{align*}
  \big| \mathcal{X} \big| << \big| \textbf{X} \big| , \big| \mathscr{Y} \big| << \big| \textbf{Y} \big|     , \\ \\   \big| \mathscr{X} \big| \approx \big| \textbf{X} \big| , \big| \mathscr{Y} \big| \approx \big| \textbf{Y} \big|        , \\ \\  \big| \mathcal{X}^{*} \big| <<  \big| \mathscr{X}^{*} \big| , \big| \mathcal{Y}^{*} \big| <<  \big| \mathscr{Y}^{*} \big|   . 
\end{align*}

\noindent The objective of choosing the alphabets, along with their pruned versions over the backward and forward conceptual channels is to increase the probabilities,

{\small \begin{align*}
  \textbf{P} \bigg[ \textit{n- bit codewords } \mathcal{C}_1 \neq \mathcal{C}_2 \textit{ transmitted at the bit transmission rate } r_2 \\ : p_{\textit{Eve false acceptance on } \mathcal{C}_1, \mathcal{C}_2} \bigg|_{\textit{Public broadcast channel}} \\ <  p_{\textit{Eve false acceptance on } \mathcal{C}_1, \mathcal{C}_2} \bigg|_{\textit{Forward conceptual channel}}  \bigg]  , 
\end{align*}}

\noindent for,

\begin{align*}
  p_{\textit{Eve commits an instance of false acceptance on } \mathcal{C}_1, \mathcal{C}_2} \equiv p_{\textit{Eve false acceptance on } \mathcal{C}_1, \mathcal{C}_2}   , 
\end{align*}

\noindent as much as possible over Eve's forward conceptual channel, while simultaneously decreasing the probabilities,

\begin{align*}
  \textbf{P} \bigg[ \textit{n- bit codewords } \mathcal{C}_1 \neq \mathcal{C}_2 \textit{ transmitted at the bit transmission rate } r_3 \\ : p_{\textit{Alice false acceptance on } \mathcal{C}_1, \mathcal{C}_2} \bigg|_{\textit{Backward conceptual channel}} \\ < p_{\textit{Eve false acceptance on } \mathcal{C}_1, \mathcal{C}_2} \bigg|_{\textit{Forward conceptual channel}}  \\ , p_{\textit{Bob false acceptance on } \mathcal{C}_1, \mathcal{C}_2} \bigg|_{\textit{Backward conceptual channel}} \\ < p_{\textit{Eve false acceptance on } \mathcal{C}_1, \mathcal{C}_2}   \bigg|_{\textit{Forward conceptual channel}}      \bigg]  , 
\end{align*}

\noindent as much as possible over Alice and Bob's backward conceptual channel, for,

\begin{align*}
  p_{\textit{Alice commits an instance of false acceptance on } \mathcal{C}_1, \mathcal{C}_2} \equiv         p_{\textit{Alice false acceptance on } \mathcal{C}_1, \mathcal{C}_2}      ,  \\   p_{\textit{Bob commits an instance of false acceptance on } \mathcal{C}_1, \mathcal{C}_2} \equiv     p_{\textit{Bob false acceptance on } \mathcal{C}_1, \mathcal{C}_2}        . 
\end{align*}

\noindent The first main result encapsulates how errors that Alice, Bob, and Eve can encounter throughout the communication protocol occur over the public broadcast, forward, and backward, conceptual channels.

\bigskip

\noindent \textbf{Theorem} \textit{1} (\textit{expressions for the adaptive bit transmission rates } $r_1, r_2, r_3$). Suppose that the noise over independent channels of Alice, Bob and Eve are $N_A$, $N_B$, and $N_E$, respectively, with corresponding error probability $\epsilon_A$, $\epsilon_B$, and $\epsilon_E$. Additionally, denote,

\begin{align*}
  \textbf{X} \bigg|_{\textit{Public broadcast channel}}  ,  \textbf{Y} \bigg|_{\textit{Public broadcast channel}} , \\   \textbf{X} \bigg|_{\textit{Eve's conceptual channel}}  ,  \textbf{Y} \bigg|_{\textit{Eve's conceptual channel}} , \end{align*}
  
  \begin{align*} \textbf{X} \bigg|_{\textit{Alice and Bob's backward conceptual channel}}  ,  \textbf{Y} \bigg|_{\textit{Alice and Bob's backward conceptual channel}} ,
\end{align*}

\noindent corresponding to the restriction of Alice and Bob's alphabets over the public broadcast, forward conceptual, and backward conceptual, channels, respectively. The three bit transmission rates are given by the expressions,

{\small \begin{align*}
   r_1 \equiv   \left\{\!\begin{array}{ll@{}>{{}}l} 
     \mathrm{log} \mathrm{log} \bigg[   \frac{  \mathrm{log} \big| \textbf{Y}^{*} \big|   }{ \big| \textbf{X}^{*} \big|  }          \bigg]   +   \mathrm{log}  \bigg[   \frac{   \mathrm{log}\big|  \textbf{Z}  \big|  }{  \big|  \textbf{Y}^{*}  \big|   } \bigg]  \Longleftrightarrow \big| \textbf{X} \big| > \big| \textbf{Y}^{*} \big| ,  \big| \textbf{Y}^{*} \big| > \big| \textbf{Z} \big|    ,     \\ \mathrm{log} \mathrm{log} \bigg[     \frac{  \mathrm{log} \big| \textbf{X} \big|  }{ \big| \textbf{Y}^{*} \big| }          \bigg]   +   \mathrm{log}  \bigg[  \frac{ \mathrm{log} \big|  \textbf{Y}^{*} \big|   }{  \big| \textbf{X} \big|   } \bigg]   \Longleftrightarrow \big| \textbf{X} \big| <  \big| \textbf{Y}^{*} \big| ,  \big|  \textbf{Y}^{*} \big| < \big| \textbf{Z} \big|        ,  \\ \mathrm{log} \mathrm{log} \bigg[     \frac{  \mathrm{log} \big| \textbf{Y}^{*} \big|  }{ \big| \textbf{X}^{*} \big| }          \bigg]   +   \mathrm{log}  \bigg[  \frac{  \mathrm{log}\big|  \textbf{Y}^{*} \big|   }{ \big| \textbf{X} \big|   } \bigg]   \Longleftrightarrow \big| \textbf{X} \big| >   \big| \textbf{Y}^{*} \big| ,  \big|  \textbf{Y}^{*} \big| < \big| \textbf{Z} \big|        ,     \\ \mathrm{log} \mathrm{log} \bigg[     \frac{  \mathrm{log} \big| \textbf{Y}^{*} \big|  }{\big| \textbf{X}^{*} \big| }          \bigg]   +   \mathrm{log}  \bigg[  \frac{  \mathrm{log}\big|  \textbf{Z}  \big|   }{ \big| \textbf{Y}^{*} \big|   } \bigg]   \Longleftrightarrow \big| \textbf{X} \big| <    \big| \textbf{Y}^{*} \big| ,  \big|  \textbf{Y}^{*} \big| >   \big| \textbf{Z} \big|        ,                          
\end{array}\right.        , \\  \\ r_2 \equiv    \left\{\!\begin{array}{ll@{}>{{}}l} 
     \mathrm{log} \mathrm{log} \bigg[   \frac{  \mathrm{log} \big| \mathcal{Y}^{*} \big|   }{ \big| \mathcal{X}^{*} \big|  }          \bigg]   +   \mathrm{log}  \bigg[   \frac{   \mathrm{log}\big|  \textbf{Z}  \big|  }{  \big|  \mathcal{Y}^{*}  \big|   } \bigg]  \Longleftrightarrow \big| \mathcal{X} \big| > \big| \mathcal{Y}^{*} \big| ,  \big| \mathcal{Y}^{*} \big| > \big| \textbf{Z} \big|    ,     \\ \mathrm{log} \mathrm{log} \bigg[     \frac{  \mathrm{log} \big| \mathcal{X} \big|  }{ \big| \mathcal{Y}^{*} \big| }          \bigg]   +   \mathrm{log}  \bigg[  \frac{ \mathrm{log} \big|  \mathcal{Y}^{*} \big|   }{  \big| \mathcal{X} \big|   } \bigg]   \Longleftrightarrow \big| \mathcal{X} \big| <  \big| \mathcal{Y}^{*} \big| ,  \big|  \mathcal{Y}^{*} \big| < \big| \textbf{Z} \big|        ,  \\ \mathrm{log} \mathrm{log} \bigg[     \frac{  \mathrm{log} \big| \mathcal{Y}^{*} \big|  }{ \big| \mathcal{X}^{*} \big| }          \bigg]   +   \mathrm{log}  \bigg[  \frac{  \mathrm{log}\big|  \mathcal{Y}^{*} \big|   }{ \big| \mathcal{X} \big|   } \bigg]   \Longleftrightarrow \big| \mathcal{X} \big| >   \big| \mathcal{Y}^{*} \big| ,  \big|  \mathcal{Y}^{*} \big| < \big| \textbf{Z} \big|        ,     \\ \mathrm{log} \mathrm{log} \bigg[     \frac{  \mathrm{log} \big| \mathcal{Y}^{*} \big|  }{\big| \mathcal{X}^{*} \big| }          \bigg]   +   \mathrm{log}  \bigg[  \frac{  \mathrm{log}\big|  \textbf{Z}  \big|   }{ \big| \mathcal{Y}^{*} \big|   } \bigg]   \Longleftrightarrow \big| \mathcal{X} \big| <    \big| \mathcal{Y}^{*} \big| ,  \big|  \mathcal{Y}^{*} \big| >   \big| \textbf{Z} \big|        ,                          
\end{array}\right.       , \\ \\ r_3 \equiv    \left\{\!\begin{array}{ll@{}>{{}}l} 
     \mathrm{log} \mathrm{log} \bigg[   \frac{  \mathrm{log} \big| \mathscr{Y}^{*} \big|   }{ \big| \mathscr{X}^{*} \big|  }          \bigg]   +   \mathrm{log}  \bigg[   \frac{   \mathrm{log}\big|  \textbf{Z}  \big|  }{  \big|  \mathscr{Y}^{*}  \big|   } \bigg]  \Longleftrightarrow \big| \mathscr{X} \big| > \big| \mathscr{Y}^{*} \big| ,  \big| \mathscr{Y}^{*} \big| > \big| \textbf{Z} \big|    ,     \\ \mathrm{log} \mathrm{log} \bigg[     \frac{  \mathrm{log} \big| \mathscr{X} \big|  }{ \big| \mathscr{Y}^{*} \big| }          \bigg]   +   \mathrm{log}  \bigg[  \frac{ \mathrm{log} \big|  \mathscr{Y}^{*} \big|   }{  \big| \mathscr{X} \big|   } \bigg]   \Longleftrightarrow \big| \mathscr{X} \big| <  \big| \mathscr{Y}^{*} \big| ,  \big|  \mathscr{Y}^{*} \big| < \big| \textbf{Z} \big|        ,  \\ \mathrm{log} \mathrm{log} \bigg[     \frac{  \mathrm{log} \big| \mathscr{Y}^{*} \big|  }{ \big| \mathscr{X}^{*} \big| }          \bigg]   +   \mathrm{log}  \bigg[  \frac{  \mathrm{log}\big|  \mathscr{Y}^{*} \big|   }{ \big| \mathscr{X} \big|   } \bigg]   \Longleftrightarrow \big| \mathscr{X} \big| >   \big| \mathscr{Y}^{*} \big| ,  \big|  \mathscr{Y}^{*} \big| < \big| \textbf{Z} \big|        ,     \\ \mathrm{log} \mathrm{log} \bigg[     \frac{  \mathrm{log} \big| \mathscr{Y}^{*} \big|  }{\big| \mathscr{X}^{*} \big| }          \bigg]   +   \mathrm{log}  \bigg[  \frac{  \mathrm{log}\big|  \textbf{Z}  \big|   }{ \big| \mathscr{Y}^{*} \big|   } \bigg]   \Longleftrightarrow \big| \mathscr{X} \big| <    \big| \mathscr{Y}^{*} \big| ,  \big|  \mathscr{Y}^{*} \big| >   \big| \textbf{Z} \big|        ,                          
\end{array}\right.        .
\end{align*}}

\noindent Besides the above result related to the \textit{adaptive} bits that are transmitted over the three Quantum channels of interest, one also must introduce error correction, and false acceptance, probabilities for the three independent channels. Determining the stochastic domination between Alice, Bob, and Eve's independent channels provides a stark contrast to how Eve loses \textit{superiority}, despite initially having it.

\bigskip

\noindent \textbf{Theorem} \textit{2} (\textit{stochastic domination of error correction, and false acceptance, probabilities over the three independent channels of Alice, Bob, and Eve}). Under the same choice of parameters provided in the previous result above, denote the probabilities,

\begin{align*}
 \underline{P_{\mathrm{EC}, A }} \equiv         \underset{A }{\underset{\mathrm{ec} \in \mathrm{EC}}{\mathrm{sup}}}    p_{\mathrm{ec}} \equiv          \mathrm{sup} \big\{ \text{success probability of a player using an error cor-} \\ \text{ recting code over Alice's independent Quantum channel} \big\}               \text{, }  \\ \\  \underline{P_{\mathrm{FA}, A }} \equiv           \underset{A }{\underset{\mathrm{fa} \in \mathrm{FA}}{\mathrm{inf}}}    p_{\mathrm{fa}} \equiv           \mathrm{inf} \big\{ \text{failure probability of a player accepting a message that} \\    \text{  should have not been accepted over Alice's independent Quantum channel} \big\}                    \text{, } \\ \\ \underline{P_{\mathrm{EC}, B }} \equiv         \underset{B }{\underset{\mathrm{ec} \in \mathrm{EC}}{\mathrm{sup}}}    p_{\mathrm{ec}} \equiv          \mathrm{sup} \big\{ \text{success probability of a player using an error correcting} \\ \text{ code over Bob's independent Quantum channel} \big\}               \text{, }  \\ \\  \underline{P_{\mathrm{FA}, B }} \equiv           \underset{B}{\underset{\mathrm{fa} \in \mathrm{FA}}{\mathrm{inf}}}    p_{\mathrm{fa}} \equiv           \mathrm{inf} \big\{ \text{failure probability of a player accepting a mes-} \\    \text{ sage that should have not been accepted over Bob's independent} \\    \text{ Quantum channel} \big\}                    \text{, }  \\ \\   \underline{P_{\mathrm{EC}, E}} \equiv         \underset{E}{\underset{\mathrm{ec} \in \mathrm{EC}}{\mathrm{sup}}}    p_{\mathrm{ec}} \equiv          \mathrm{sup} \big\{ \text{success probability of a player using an error cor} \\    \text{ -recting code over Eve's independent Quantum channel} \big\}               \text{, }  \\ \\  \underline{P_{\mathrm{FA}, E}} \equiv           \underset{E}{\underset{\mathrm{fa} \in \mathrm{FA}}{\mathrm{inf}}}    p_{\mathrm{fa}} \equiv           \mathrm{inf} \big\{ \text{failure probability of a player accepting a message that} \\ \text{ should  have not been accepted over Eve's independent Quantum channel} \big\}                    \text{, }
\end{align*}

\noindent along with the following restrictions of the above defined probabilities over the public broadcast, forward conceptual, and backward conceptual, channels,

\begin{align*}
  \underline{P_{\mathrm{EC}, A }}  \bigg|_{\textit{Public broadcast channel}}   ,   \underline{P_{\mathrm{EC}, B }}  \bigg|_{\textit{Public broadcast channel}} ,   \underline{P_{\mathrm{EC}, E }}  \bigg|_{\textit{Public broadcast channel}}  ,   \\   \underline{P_{\mathrm{FA}, A }}  \bigg|_{\textit{Public broadcast channel}}   ,   \underline{P_{\mathrm{FA}, B }}  \bigg|_{\textit{Public broadcast channel}} ,   \underline{P_{\mathrm{FA}, E }}  \bigg|_{\textit{Public broadcast channel}}  ,   \\  \\   \underline{P_{\mathrm{EC}, A }}  \bigg|_{\textit{Eve's conceptual channel}}   ,   \underline{P_{\mathrm{EC}, B }}  \bigg|_{\textit{Eve's conceptual channel}} ,   \underline{P_{\mathrm{EC}, E }}  \bigg|_{\textit{Eve's conceptual channel}}  ,   \\   \underline{P_{\mathrm{FA}, A }}  \bigg|_{\textit{Eve's conceptual channel}}   ,   \underline{P_{\mathrm{FA}, B }}  \bigg|_{\textit{Eve's conceptual channel}} ,   \underline{P_{\mathrm{FA}, E }}  \bigg|_{\textit{Eve's conceptual channel}}  ,   \\   \\   \underline{P_{\mathrm{EC}, A }}  \bigg|_{\textit{Alice and Bob's backward conceptual channel}}   ,   \underline{P_{\mathrm{EC}, B }}  \bigg|_{\textit{Alice and Bob's backward conceptual channel}} , \\   \underline{P_{\mathrm{EC}, E }}  \bigg|_{\textit{Alice and Bob's backward conceptual channel}}  ,   \\   \underline{P_{\mathrm{FA}, A }}  \bigg|_{\textit{Alice and Bob's backward conceptual channel}}   ,   \underline{P_{\mathrm{FA}, B }}  \bigg|_{\textit{Alice and Bob's backward conceptual channel}} , \\   \underline{P_{\mathrm{FA}, E }}  \bigg|_{\textit{Alice and Bob's backward conceptual channel}}  ,       
\end{align*}

\noindent where,

\begin{align*}
 \mathrm{EC} \equiv \underset{A,B,E}{\underset{\mathrm{ec} \in \mathrm{EC}}{\bigcup}}  \big\{ \textit{error correcting codes } \mathrm{ec} \big\}  , \\ \\    \mathrm{FA} \equiv   \underset{A,B,E}{\underset{\mathrm{fa} \in \mathrm{FA}}{\bigcup}}  \big\{ \textit{instances of false acceptance } \mathrm{fa} \big\}       .
\end{align*}

\noindent One has that,

{\small \[ \left\{\!\begin{array}{ll@{}>{{}}l}    
 \underline{P_{\mathrm{FA}, E }}  \bigg|_{\textit{Public broadcast channel}} <  \underline{P_{\mathrm{FA}, A }}  \bigg|_{\textit{Public broadcast channel}}   ,  \underline{P_{\mathrm{FA}, E }}  \bigg|_{\textit{Public broadcast channel}} \\ <  \underline{P_{\mathrm{FA}, B }}  \bigg|_{\textit{Public broadcast channel}} , \\  \underline{P_{\mathrm{EC}, E }}  \bigg|_{\textit{Public broadcast channel}} >  \underline{P_{\mathrm{EC}, A }}  \bigg|_{\textit{Public broadcast channel}}   ,  \underline{P_{\mathrm{EC}, E }}  \bigg|_{\textit{Public broadcast channel}} \\ >  \underline{P_{\mathrm{EC}, B }}  \bigg|_{\textit{Public broadcast channel}} ,  \tag{1} \end{array}\right.
\]

\bigskip

\[ \left\{\!\begin{array}{ll@{}>{{}}l}   \underline{P_{\mathrm{FA}, E }}  \bigg|_{\textit{Eve's conceptual channel}}  >     \underline{P_{\mathrm{FA}, A }}  \bigg|_{\textit{Eve's conceptual channel}}         ,  \underline{P_{\mathrm{FA}, E }}  \bigg|_{\textit{Eve's conceptual channel}}  \\ >     \underline{P_{\mathrm{FA}, B }}  \bigg|_{\textit{Eve's conceptual channel}} ,  \\      \underline{P_{\mathrm{EC}, E }}  \bigg|_{\textit{Eve's conceptual channel}} <  \underline{P_{\mathrm{EC}, A }}  \bigg|_{\textit{Eve's conceptual channel}}   ,  \underline{P_{\mathrm{EC}, E }}  \bigg|_{\textit{Eve's conceptual channel}} \\ <   \underline{P_{\mathrm{EC}, B }}  \bigg|_{\textit{Eve's conceptual channel}} ,  \tag{2} \end{array}\right. 
\] 

\bigskip

\[ \left\{\!\begin{array}{ll@{}>{{}}l}      \underline{P_{\mathrm{FA}, E }}  \bigg|_{\textit{Public broadcast channel}} > \underline{P_{\mathrm{FA}, A }}  \bigg|_{\textit{Alice and Bob's backward conceptual channel}}  ,     \underline{P_{\mathrm{FA}, E }}  \bigg|_{\textit{Public broadcast channel}} \\ > \underline{P_{\mathrm{FA}, B }}  \bigg|_{\textit{Alice and Bob's backward conceptual channel}} , \\     \underline{P_{\mathrm{EC}, E }}  \bigg|_{\textit{Public broadcast channel}} < \underline{P_{\mathrm{EC}, A }}  \bigg|_{\textit{Alice and Bob's backward conceptual channel}}  ,     \underline{P_{\mathrm{EC}, E }}  \bigg|_{\textit{Public broadcast channel}} \\ < \underline{P_{\mathrm{EC}, B }}  \bigg|_{\textit{Alice and Bob's backward conceptual channel}}              .  \tag{3}
\end{array}\right.
\] }

\noindent The following properties of $p_{\mathrm{FA},\mathrm{E}}$ hold:

\begin{itemize}

 \item [$\bullet$] \textit{Minimization of probabilities via the Fano inequality}. One has that,

\begin{align*}
  p_{\mathrm{FA},\mathrm{E}} \equiv \underset{\chi_E > 0}{\underset{M > 0}{\bigcup}}  \mathrm{inf} \big\{ p :      \mathrm{log} M - \chi_E \leq  H_b \big(  p \big)   + p \mathrm{log}  \big[ M - 1 \big]          \big\}   .
\end{align*}

\item[$\bullet$] \textit{Up to constant lowers bounds for Eve's error probability}. One has that,

\begin{align*}
  P_{\mathrm{FA},\mathrm{E}} \gtrsim  \mathrm{log} \bigg[ \frac{1}{N} \bigg] \bigg[ \mathrm{log} M - \chi_E - 1 \bigg]  ,
\end{align*}

\noindent for $N \neq M$, each of which can be taken to be sufficiently large.

\bigskip

\item[$\bullet$] \textit{Sharpening the up to constants lower bound for Eve's error probability to an inequality via Helstrom}. One has that,

\begin{align*}
    P_{\mathrm{FA}, \mathrm{E}} \geq 1 - \frac{1 + \epsilon \big( M - 1 \big)}{M}      , 
\end{align*}

\noindent for $\epsilon$ taken to be sufficiently small.

\end{itemize}

\bigskip

\noindent Finally, the remaining main result below provides conditions under which Alice and Bob can map into their shared authenticated space with high probability.

\bigskip

\noindent \textbf{Theorem} \textit{3} (\textit{Alice and Bob can map into their shared authenticated space of bit codewords with high probability. Their authentication probability over the backward conceptual channel with secrecy capacity}, $C_s$, \textit{ stochastically dominates the authentication probability over the public broadcast channel}.) Fix $0 \leq p < q \leq \frac{1}{2}$. For $r_1, r_2, r_3 > h \big( q \big) - h \big( p \big)$, and the set of $n$-bit transmission mechanisms,

\begin{align*}
    \mathcal{N}^{n}_{p,q} \equiv    \mathcal{N}^{n}_{p,q} \bigg|_{\textit{Public broadcast channel}} \overset{\cdot}{\cup}   \mathcal{N}^{n}_{p,q} \bigg|_{\textit{Alice and Bob's backward conceptual channel}}   , 
\end{align*}

\noindent introduce the probabilities,

\begin{align*}
  P_{\textit{Mapping into the authenticated space of the public broadcast channel}}  , \\ \\ P_{\textit{Mapping into the authenticated space of Eve's forward conceptual channel}} , \end{align*}

  \begin{align*} P_{\textit{Mapping into the authenticated space of Alice and Bob's backward conceptual channel}} , 
\end{align*}

\noindent which can be respectively decomposed as,

\begin{align*}
 \textbf{P} \big[ \textit{encodings, decodings, bit transmission rates} :  \textit{Alice and Bob map into the authenticated space} \\ \textit{ of the public broadcast channel at a specified bit transmission rate} \big] , \end{align*}

 \begin{align*}  \textbf{P} \big[ \textit{encodings, decodings, bit transmission rates} :  \textit{Eve maps into the authenticated space} \\ \textit{ of the public broadcast channel at a specified bit transmission rate} \big], \end{align*}

 \begin{align*}
 \textbf{P} \big[ \textit{encodings, decodings, bit transmission rates} :  \textit{Alice and Bob map into the authenticated space} \\ \textit{ of the conceptual backward channel at a specified bit transmission rate} \big] , 
\end{align*}

\noindent and protocols,

\begin{align*}
   \pi^n_1 \equiv   \pi^n \bigg|_{\textit{Public broadcast channel}}  \equiv  \bigg[  E^n   \bigg|_{\textit{Public broadcast channel}}   ,  D^n \bigg|_{\textit{Public broadcast channel}} \bigg] \\ \\  \equiv          \big[ \textit{Alice's protocol for decoding n-bit codewords} \\  , \textit{Bob's protocol for decoding n-bit codewords}  \big] \big|_{\textit{Public broadcast channel}}   , \\ \\    \pi^n_2 \equiv    \pi^n \bigg|_{\textit{Backward conceptual channel}}  \equiv  \bigg[   E^n \bigg|_{\textit{Backward conceptual channel}}     ,  D^n \bigg|_{\textit{Backward conceptual channel}} \bigg]  \\ \\  \equiv         \big[ \textit{Alice's protocol for decoding n-bit codewords} \\ , \textit{Bob's protocol for decoding n-bit codewords}  \big] \big|_{\textit{Backward conceptual channel}}  , 
\end{align*}

\noindent over $n$ bit codewords so that,

\begin{align*}
\mathcal{N}^n_{1,p,q} \equiv  \mathcal{N}^n_{p,q} \bigg|_{\textit{Public broadcast channel}} \overset{\pi^n_1}{\longrightarrow} \mathcal{A}^n_{1,p,q} \equiv \mathcal{A}^n_{p,q} \bigg|_{\textit{Public broadcast}}  , \\ \\   \mathcal{N}^n_{2,p,q} \equiv  \mathcal{N}^n_{p,q} \bigg|_{\textit{Alice and Bob's backward conceptual channels}} \overset{\pi^n_2}{\longrightarrow} \mathcal{A}^n_{1,p,q} \\ \equiv \mathcal{A}^n_{p,q} \bigg|_{\textit{Alice and Bob's backward conceptual channels}} .  
\end{align*}

\noindent \textbf{Corollary} \textit{1} (\textit{stability of Alice and Bob's protocols for mapping into the authenticated space for codewords with infinitely many bits}). Fix the same choice of parameters provided in \textbf{Theorem} \textit{3}. For transmitted codewords with infinitely many bits, the protocols,

\begin{align*}
 \underset{n \longrightarrow + \infty}{\mathrm{lim}}  \pi^n_1 \equiv   \underset{n \longrightarrow + \infty}{\mathrm{lim}}   \pi^n \bigg|_{\textit{Public broadcast channel}}  \equiv   \underset{n \longrightarrow + \infty}{\mathrm{lim}}  \bigg[  E^n   \bigg|_{\textit{Public broadcast channel}}  \\ ,  D^n \bigg|_{\textit{Public broadcast channel}} \bigg] \\ \\   \underset{n \longrightarrow + \infty}{\mathrm{lim}}   \pi^n_2 \equiv   \underset{n \longrightarrow + \infty}{\mathrm{lim}}   \pi^n \bigg|_{\textit{Backward conceptual channel}}  \equiv   \underset{n \longrightarrow + \infty}{\mathrm{lim}}  \bigg[   E^n \bigg|_{\textit{Backward conceptual channel}}   \\  ,  D^n \bigg|_{\textit{Backward conceptual channel}} \bigg]   , 
\end{align*}

\noindent of Alice and Bob's protocols, respectively, exist and guarantee that bit codewords with infinitely many bits can be mapped into the authenticated space whp. Explicitly, such mappings are  given by,

\begin{align*}
 \underset{n \longrightarrow + \infty}{\mathrm{lim}} \mathcal{N}^n_{1,p,q} \equiv   \underset{n \longrightarrow + \infty}{\mathrm{lim}} \mathcal{N}^n_{p,q} \bigg|_{\textit{Public broadcast channel}} \overset{\pi^{+\infty}_1}{\longrightarrow}  \underset{n \longrightarrow + \infty}{\mathrm{lim}} \mathcal{A}^n_{1,p,q} \equiv  \underset{n \longrightarrow + \infty}{\mathrm{lim}} \mathcal{A}^n_{p,q} \bigg|_{\textit{Public broadcast}}  , \\ \\  \\    \underset{n \longrightarrow + \infty}{\mathrm{lim}} \mathcal{N}^n_{2,p,q} \equiv   \underset{n \longrightarrow + \infty}{\mathrm{lim}} \mathcal{N}^n_{p,q} \bigg|_{\textit{Alice and Bob's backward conceptual channels}} \overset{\pi^{+\infty}_2}{\longrightarrow}  \underset{n \longrightarrow + \infty}{\mathrm{lim}} \mathcal{A}^n_{1,p,q} \\ \equiv  \underset{n \longrightarrow + \infty}{\mathrm{lim}} \mathcal{A}^n_{p,q} \bigg|_{\textit{Alice and Bob's backward conceptual channels}} .  
\end{align*}

\noindent Associated with the two above protocols, are the objects,

\begin{align*}
   \underset{n \longrightarrow + \infty}{\mathrm{lim}}  \mathcal{N}^{n}_{p,q} \equiv   \underset{n \longrightarrow + \infty}{\mathrm{lim}} \bigg\{   \mathcal{N}^{n}_{p,q} \bigg|_{\textit{Public broadcast channel}} \overset{\cdot}{\cup}   \mathcal{N}^{n}_{p,q} \bigg|_{\textit{Alice and Bob's backward conceptual channel}} \bigg\}   , 
\end{align*}

\noindent corresponding to the collection of transmission mechanisms, for codewords with infinitely many bits, over Alice and Bob's shared channel.

\bigskip

\noindent \textbf{Corollary} \textit{2} (\textit{the expected gap of Eve's false acceptance probability over the main, and forward conceptual, Quantum channels}). One has that,

\begin{align*}C_{\mathrm{Eve}} \approx 1 - \frac{1 + \epsilon \big( M - 1 \big)}{M}      , 
\end{align*} 

\noindent for small $M$, and otherwise,

\begin{align*}C_{\mathrm{Eve}} \backsimeq     \mathrm{log} \bigg[ \frac{1}{N} \bigg] \bigg[ \mathrm{log} M - \chi_E - 1 \bigg]                 , 
\end{align*}

\noindent from the expression obtained for (Eve's false acceptance probability gap).

\bigskip

\noindent In the next section, we provide arguments for each of the main results.

\section{Arguments for the main results}

\subsection{\textbf{Theorem} \textit{1}}

\noindent \textit{Proof of Theorem 1}. Apply previous arguments of the author provided in {\color{blue}[48]}, specifically in \textbf{Theorem} \textit{1}, to obtain the desired bit transmission rates, $r_1, r_2. r_3$, from which we conclude the argument. \boxed{}

\subsection{\textbf{Theorem} \textit{2}}

\noindent \textit{Proof of Theorem 2}. Fix $M>0$, and $\mathrm{log} \big[ \cdot \big]$ as the base two natural logarithm. Denote $\chi_E$ as Eve's Holevo information,

 \begin{align*}
    \chi_E \equiv  S \big[ \bar{\sigma} \big] - \frac{1}{M} \underset{u \in \mathcal{H}}{\sum} S \big[ \sigma_u \big]      ,
 \end{align*}

\noindent $H_b$ as the binary entropy function,

\begin{align*}
 H_b \big( \cdot \big) \equiv \cdot \times \mathrm{log} \big[ \cdot \big] +  \big( 1 - \cdot \big)  \times \mathrm{log} \big[ 1 - \cdot \big]  ,
\end{align*}

\noindent in addition to the function,

\begin{align*}
   f \big( p \big) \equiv H_b \big[ p \big] + p \text{ }  \mathrm{log} \big[ M - 1 \big] ,
\end{align*}

\noindent for the probability $p$, where $0 < p < 1 $.

\bigskip

\noindent It suffices to demonstrate that Eve's error probability $P_{\mathrm{FA},\mathrm{E}}$ implicitly takes the form,

\begin{align*}
     \mathrm{log} M - \chi_E \leq  H_b \big(  P_{\mathrm{FA},\mathrm{E}} \big)   +  P_{\mathrm{FA},\mathrm{E}} \mathrm{log}  \big[ M - 1 \big] .
\end{align*}

\noindent An inequality of the above form is expected to hold from the fact that the Mutual-Information entropy is related to the differences of the sum rates, {\color{blue}[53]},

\begin{align*}
     R \big( Q \big) \geq R \big( W \big) - \delta \big( \epsilon \big)      ,
\end{align*}

\noindent for a channel degradation of $Q$, with $W$, with,

\begin{align*}
  \delta \big( \epsilon \big) \longrightarrow 0  ,
\end{align*}

\noindent as the fidelity $\epsilon \longrightarrow 0^{+}$. The polarization construction of codes over $Q$ and $W$, besides having linear-time complexity, implies, within the current setting, that the Mutual Information entropy over the two degraded channels considered in the current setting satisfies,

\begin{align*}
     I \big( Q \big) \geq I \big( W \big) - \delta \big( \epsilon \big)      ,
\end{align*}

\noindent under the same assumption on the rate of decay of the fidelity. In terms of quantities related to Holevo, the above relation corresponding to polarization defined for two degraded channels takes the form,

\begin{align*}
   I \big( W \big) \equiv   \chi \big\{ \rho_x , \rho_x \big\} = S \bigg[ \frac{1}{\big| \mathcal{X} \big|}   \underset{x \in \mathcal{X}}{\sum} \rho_x                    \bigg] -  \frac{1}{\big| \mathcal{X} \big|}   S \bigg[      \underset{x \in \mathcal{X}}{\sum} \rho_x      \bigg]      ,
\end{align*}

\noindent for,

\begin{align*}
\big|   \mathcal{X}^0  \big|  \equiv \big|   \mathcal{X} \big| \equiv \textit{Size of the alphabet corresponding to the first use of a channel } \mathcal{X}    , \\ \vdots  \\ \big|   \mathcal{X}^t \big| \equiv \textit{Size of the alphabet corresponding to the $t>0$ use of a channel } \mathcal{X}    , \\ \\     \big|   \mathcal{X}^0  \big|^{-1}  \equiv \big|   \mathcal{X} \big|^{-1} \equiv \textit{Normalization  of the alphabet corresponding to the first use of a channel } \mathcal{X}    , \\ \vdots  \\ \big|   \mathcal{X}^t \big|^{-1} \equiv \textit{Normalization of the alphabet corresponding to the $t>0$ use of a channel } \mathcal{X}      ,   \\ \\ \rho_x \equiv \textit{Quantum state corresponding to the value output by the Quantum multiple-access} \\ \textit{channel x}     , \\ \\   \rho_{x_1 x_2} \equiv  \underset{x_1 \leq i \leq x_2}{\prod} \rho_i \equiv \textit{Quantum state corresponding to the value output by the Quan-} \\ \textit{tum multiple-access channel $x_1,x_2$}     ,    \\ \vdots \\   \rho_{x_1 \cdots x_N}  \equiv  \underset{x_1 \leq i \leq x_N}{\prod} \rho_i \equiv \textit{Quantum state corresponding to the value output by the Quan-} \\ \textit{tum multiple-access channel $x_1 \cdots x_N$}          , 
\end{align*}

\noindent and the channel mapping,

\begin{align*}
    W : x \in \mathcal{X} \mapsto \rho_x \in \mathcal{D} \big( \mathcal{H} \big) \equiv \big\{ \textit{Distribution space over the Hilbert space $\mathcal{H}$} \big\} \\ \equiv \underset{\textit{generators of an instance of $\mathcal{D}$}}{\bigcup}  \bigg\{ \textit{generators: generators have support over $\mathcal{H}$} \bigg\}      , 
\end{align*}

\noindent corresponding to $W$.

\bigskip

\noindent With the above expression for the symmetric information entropy, in the setting of this paper, one computes the conserved quantity,

\begin{align*}
  \frac{1}{2}  \bigg\{   S \bigg[ \frac{1}{\big| \mathcal{X}^+ \big|}   \underset{x \in \mathcal{X}^+}{\sum} p^+_x                    \bigg] -  \frac{1}{\big| \mathcal{X}^+ \big|}   S \bigg[      \underset{x \in \mathcal{X}^+}{\sum} p^+_x      \bigg]    +  S \bigg[ \frac{1}{\big| \mathcal{X}^- \big|}   \underset{x \in \mathcal{X}^-}{\sum} p^-_x                    \bigg] -  \frac{1}{\big| \mathcal{X}^- \big|}   S \bigg[      \underset{x \in \mathcal{X}^-}{\sum} p^-_x      \bigg]    \bigg\} 
\end{align*}

\noindent of Holevo sums corresponding to $I \big( W \big)$, where $p^{\pm}$ denote the same quantites as presented above, under the notation that an original channel $W$ can be used to produce the superior, and inferior, channels $W^+$ and $W^-$, respectively.

\bigskip

\noindent In the below items, we demonstrate how the Completely Positive Trace Preserving (CPTP) map, specifically a map of the form,

\begin{align*}
  \Phi : \mathcal{B } \big[ \mathcal{H}_A \big] \longrightarrow \mathcal{B} \big[ \mathcal{H}_B \big]  ,
\end{align*}

\noindent for the preimage, and image,

\begin{align*}
  \mathcal{B } \big[ \mathcal{H}_A \big]  \equiv \textit{Set of channel outputs with support over $\mathcal{H}_A$}  , \\ \\ \mathcal{B } \big[ \mathcal{H}_B \big] \equiv  \textit{Set of channel outputs with support over $\mathcal{H}_B$} , 
\end{align*}

\noindent are used to measure the degraded channel from the original channel. Such maps satisfy:

\begin{itemize}
    \item[$\bullet$] \textit{CPTP positivity}. The image of the CPTP map into $\mathcal{B} \big[ \mathcal{H}_A \big]$ is strictly positive iff the input probability itself is strictly positive.

\bigskip

\item[$\bullet$] \textit{CPTP complete positivity}. A CPTP map is completely positive iff the tensor product of the CPTP map with the identity operator is positive.

    \noindent

    \bigskip

     \item[$\bullet$] \textit{CPTP maps are linear}. One has that,

     \begin{align*}
  \Phi \big[ \alpha \rho + \beta \sigma  \big] =  \alpha \Phi \big[ \rho \big]  + \beta  \Phi \big[  \sigma  \big]                ,
     \end{align*}

     \noindent for $\alpha, \beta \in \textbf{C}$, and,

       \begin{align*}
      \rho \equiv  \textit{State distributed to Eve with support over $\mathcal{H}_A $}  , \\ \\ \sigma \equiv \textit{State distributed to Eve with support over $\mathcal{H}_B $} . 
    \end{align*}

     \bigskip

      \item[$\bullet$] \textit{CPTP tensor product closure}. Denote $\Phi^{\prime}$ as another CPTP which is not equal to $\Phi$. Then $\Phi^{\prime} \otimes \Phi$ is also CPTP.

\bigskip

      \item[$\bullet$] \textit{CPTP composition closure}. Denote $\Phi^{\prime}$ as another CPTP which is not equal to $\Phi$. Then the composition $\Phi^{\prime} \circ \Phi$ is also CPTP.

      \bigskip

       \item[$\bullet$] \textit{CPTP contractivity with respect to the trace norm}. Eve's error probability is shown to satisfy the above inequality with respect to the binary entropy function as a result of the \textit{CPTP contractivity property}, which states,

       \begin{align*}
        \big| \big| \Phi \big[ \rho \big] - \Phi \big[ \sigma \big] \big| \big|_1 \leq \big| \big| \rho - \sigma \big| \big|_1    .
       \end{align*}

\noindent Hence,

\begin{align*}
  P_{\mathrm{FA},\mathrm{E}} \gtrsim  \mathrm{log} \bigg[ \frac{1}{N} \bigg] \bigg[ \mathrm{log} M - \chi_E - 1 \bigg]  .
\end{align*}

       \noindent as provided in \textbf{Lemma} \textit{3.2.3} below, for,

       \begin{align*}
         \rho_0 \equiv  \textit{State distributed to Eve with support over $\mathcal{H}_A $}  , \\ \\ \rho_1 \equiv \textit{State distributed to Eve with support over $\mathcal{H}_B $} . 
       \end{align*}

    \item[$\bullet$] \textit{CPTP data-processing inequality}. One has that,

    \begin{align*}
       D \big[ \rho \big| \big| \sigma \big] \geq   D \big[ \Phi \big[ \rho \big] \big| \big| \Phi \big[ \sigma \big] \big]  ,
    \end{align*}

    \noindent for,

    \begin{align*}
      \rho \equiv  \textit{State distributed to Eve with support over $\mathcal{H}_A $}  , \\ \\ \sigma \equiv \textit{State distributed to Eve with support over $\mathcal{H}_B $} . 
    \end{align*}

\end{itemize}

\noindent We make use of the above properties to obtain the desired inequality for $P_{\mathrm{FA},\mathrm{E}}$. We first demonstrate, from a suitable POVM representation, that Alice and Bob can have false acceptance probabilities that are far less than those of Eve. Besides demonstrating that the expected stochastic domination holds, we then provide several steps for concluding that Eve's false acceptance probability satisfies the desired inequality.

\bigskip

\noindent \textbf{Lemma} \textit{3.2.1} (\textit{Post-processing monotonicity}). Let $\big\{ p_u \big\}_{u \in U} $ be an ensemble of Bob's systems with prior $p_U \equiv \underset{u \in U }{\bigcup} p_u$. For a CPTP map $\mathcal{E}$, so that $\mathcal{E} \big[ p_u \big] = \sigma_u$, and POVM $\big\{ F_u \big\}$ over Eve's systems,

\begin{align*}
  \underset{F}{\mathrm{Pr}} \bigg[     \big\{  \hat{U} = u \big\}  \big|   \big\{ U = u \big\}        \bigg]  = \mathrm{Tr} \big[ F_u \sigma_u \big]    = \mathrm{Tr} \big[  \mathcal{E}^{\dagger} \big[ F_u \big] p_u  \big]         .
\end{align*}

\noindent Consequently, the success probabilities of Bob and Eve accepting a message over the channel satisfy,

\begin{align*}
   P_{\mathrm{Success}, \mathrm{Bob}}  \geq  P_{\mathrm{Success}, \mathrm{Eve}}     ,   
\end{align*}

\noindent and hence,

\begin{align*}
   P_{\mathrm{FA}, \mathrm{Eve}}  \geq  P_{\mathrm{FA}, \mathrm{Bob}}     .   
\end{align*}

\noindent Similarly,

\begin{align*}
   P_{\mathrm{FA}, \mathrm{Eve}}  \geq  P_{\mathrm{FA}, \mathrm{Alice}}     .   
\end{align*}

\noindent \textit{Proof of Lemma 3.2.1}. The adjoint map $\mathcal{E}^{\dagger}$ of $\mathcal{E}$ is $\mathcal{E}^{\dagger} \big[ \textbf{I} \big] = \textbf{I}$, and hence unital and completely positive. For the POVM $\big\{ F_u \big\}$ mentioned in the above statement,

\begin{align*}
 \underset{u \in U}{\sum} \mathcal{E}^{\dagger} \big[   F_u  \big]  = \textbf{I}    , 
\end{align*}

\noindent under the assumption that the POVM satisfies the condition $0 <  F \leq \textbf{I}$. Therefore every measurement that Eve performs corresponds to a measurement that Bob performs. Moreover, from the fact that,

\begin{align*}
  \mathrm{Tr} \big[ F_u \sigma_u \big]    = \mathrm{Tr} \big[  \mathcal{E}^{\dagger} \big[ F_u \big] p_u  \big]       ,
\end{align*}

\noindent taking the supremum over all $u$, through,

\begin{align*}
\underset{\mathrm{POVM } \text{ }  F}{\mathrm{sup}}  \big\{ \mathrm{sup}_u \big\{  \mathrm{Tr} \big[ F_u \sigma_u \big]  \big\} \big\}   =  \underset{\mathrm{POVM } \text{ }  F}{\mathrm{sup}}  \big\{ \mathrm{sup}_u  \big\{   \mathrm{Tr} \big[  \mathcal{E}^{\dagger} \big[ F_u \big] p_u  \big]   \big\} \big\}     ,
\end{align*}

\bigskip implies that any optimal value that Bob obtains is at least as much as Eve's, from which we conclude the argument. \boxed{}

\bigskip

\noindent To exhibit that Eve's error probability satisfies,

\begin{align*}
     \mathrm{log} M - \chi_E \leq  H_b \big(  P_{\mathrm{FA},\mathrm{E}} \big)   +  P_{\mathrm{FA},\mathrm{E}} \mathrm{log}  \big[ M - 1 \big] ,
\end{align*}

\noindent introduce:

\begin{itemize}
    \item[$\bullet$] \textit{Blocklength}. Write,

    \begin{align*}
     N \equiv 2^n    ,
    \end{align*}

    \noindent corresponding to the blocklength, where the inputs over each channel are binary (ie, taking values in $\big\{ 0 ,1 \big\}$).

    \bigskip

    \item[$\bullet$] \textit{Uniform prior over messages}. Write,

    \begin{align*}
       \underset{U}{\textbf{P}}  \big[ \cdot \big] ,
    \end{align*}

    \noindent corresponding to the probability measure over messages $U$, which is equivalent to the collection of all $K$-bit messages.

    \bigskip

    \item[$\bullet$] \textit{Classical-Quantum states}. Write,

    \begin{align*}
    p_u = \mathcal{E}_{\mathrm{C+PB}} \big[ \sigma_u \big]    ,
    \end{align*}

    \noindent corresponding to the Classical-Quantum states, for the CPTP map $\mathcal{E}_{\mathrm{C+PB}}$ over the cascade and public broadcast channel.

    \bigskip

    \item[$\bullet$] \textit{Averaged states}. Write,

    \begin{align*}
    \bar{p}  \equiv  \frac{1}{M} \underset{u \in U}{\sum} p_u      , \\ \\ \bar{\sigma} \equiv  \frac{1}{M} \underset{u \in U}{\sum} \sigma_u            , 
    \end{align*}

\noindent corresponding to the averaged states of $p$, and of $\sigma$, respectively, over all $u \in U$.
    
    \bigskip

    \item[$\bullet$] \textit{Fidelity parameters for binary Classical-Quantum channels}. Write,

\begin{align*}
      I \big[ W_A \big] \equiv   \textit{Symmetric information given Alice's single-use of the Classical Quantum channel}  , \\ \\    I \big[ W_B \big] \equiv \textit{Symmetric information given Bob's single-use of the Classical Quantum channel}   , \\  \\      I \big[ W_E \big]   \equiv  \textit{Symmetric information given Eve's single-use of the Classical Quantum channel}     . 
\end{align*}

    \noindent Given binary inputs over $\big\{ 0 ,1 \big\}$, $0 < I \big[ \cdot \big] < 1$.

\end{itemize}

\noindent We state the next results which are used to obtain the desired inequality for $P_{\mathrm{FA},E}$.

\bigskip

\noindent \textbf{Lemma} \textit{3.2.2} (\textit{Relating Eve's Holevo information to her error probability over the forward cascade of the public broadcast channel}). One has that,

\begin{align*}
     \mathrm{log} M - \chi_E \leq H_b \big( P_{\mathrm{FA},\mathrm{E}} \big) -   P_{\mathrm{FA},\mathrm{E}}  \mathrm{log} \big[ M - 1 \big]    .
\end{align*}

\noindent \textit{Proof of Lemma 3.2.2}. Observe,

\begin{align*}
   I \big[ U ; Z \big]   \leq \chi_E ,
\end{align*}

\noindent namely that the mutual information entropy between the set of measurements that Bob gathers, $U$, and the classical outcomes $Z$ that Eve observes, is always upper bounded by Eve's Holevo information. By Fano's inequality, one has that,

\begin{align*}
  H \big[ U \big| Z \big] \leq H_b \big( P_{\mathrm{FA}, \mathrm{E}} \big) +  P_{\mathrm{FA}, \mathrm{E}}  \mathrm{log} \big[ M - 1 \big] ,
\end{align*}

\noindent which in turn implies that, for $P_{\mathrm{FA},\mathrm{E}} \equiv \underset{U}{\textbf{P}} \big[ \hat{U} \neq U \big]$, that,

\begin{align*}
   I \big( U ; Z \big) = H \big( U \big) - H \big( U \big| Z \big) \leq \mathrm{log} M - H_b \big(  P_{\mathrm{FA}, \mathrm{E}} \big) -  P_{\mathrm{FA}, \mathrm{E}}  \mathrm{log} \big[ M - 1 \big]  . 
\end{align*}

\noindent Finally, given the first bound relating Eve's Holevo Information to the mutual information entropy, one has that,

\begin{align*}
     \mathrm{log} M - \chi_E \leq H_b \big( P_{\mathrm{FA},\mathrm{E}} \big) -   P_{\mathrm{FA},\mathrm{E}}  \mathrm{log} \big[ M - 1 \big]    ,
\end{align*}

\noindent from which we conclude the argument. \boxed{}

\bigskip

\noindent \textbf{Lemma} \textit{3.2.3} (\textit{obtaining the desired inequality for Eve's error probability over the forward cascade of the public broadcast channel from the previous result}). One has that,

\begin{align*}
  P_{\mathrm{FA},\mathrm{E}} \gtrsim  \mathrm{log} \bigg[ \frac{1}{N} \bigg] \bigg[ \mathrm{log} M - \chi_E - 1 \bigg]  .
\end{align*}

\noindent \textit{Proof of Lemma 3.2.3}. For $M$ taken large enough, under the assumption,

\begin{align*}
    H_b \big(   P_{\mathrm{FA},\mathrm{E}} \big) +   P_{\mathrm{FA},\mathrm{E}} \mathrm{log} \big[ M - 1 \big] \leq   P_{\mathrm{FA},\mathrm{E}} \mathrm{log} M + 1 
\end{align*}

\noindent because $H_b \big( p_{\mathrm{FA},\mathrm{E}} \big) < 1$, one straighforwardly obtains the desired up to constants lower bound for Eve's error probability from direct rearrangement, as we conclude the argument. \boxed{}

\bigskip

\noindent The two above results demonstrate that the first desired property of Eve's error probability provided in the statement of \textbf{Theorem} \textit{2} holds. Below we demonstrate how the up to constants lower bound for her error probability can be sharpened.

\bigskip

\noindent \textbf{Lemma} \textit{3.2.4} (\textit{sharpening the up to constants to lower bound to a lower bound independent of the constant}). One has that,

\begin{align*}
    P_{\mathrm{FA}, \mathrm{E}} \geq 1 - \frac{1 + \epsilon \big( M - 1 \big)}{M}      , 
\end{align*}

\noindent for $\epsilon$ taken to be sufficiently small.

\bigskip

\noindent \textit{Proof of Lemma 3.2.4}. Fix $\epsilon$ sufficiently small, and, as stated previously, $NM>0$. Observe that the desired constant takes the form,

\begin{align*}
 \big\{   C_H \big[ 1,  1 \big] \propto    \underset{u \neq u^{\prime}}{\mathrm{sup}} \big| \big| \sigma_u - \sigma_{u^{\prime}} \big| \big|_1  \big\}   \Longleftrightarrow  \big\{      C_H \big[ 1,  1 \big] \approx     \underset{u \neq u^{\prime}}{\mathrm{sup}} \big| \big| \sigma_u - \sigma_{u^{\prime}} \big| \big|_1  \leq \epsilon  \big\}    , 
\end{align*}

\noindent where,

\begin{align*}
   \frac{\mathrm{log} \big[ \frac{1}{N} \big] \big[ \mathrm{log} M - \chi_E - 1 \big]}{1 - \frac{1 + \epsilon \big( M - 1 \big)}{M}}    \approx    - \frac{ \bigg[       1 + \frac{1}{N} + \mathcal{O} \big[ N^2 \big]     \bigg]  \big[ \mathrm{log} M - \chi_E - 1 \big]}{1 - \frac{1 + \epsilon \big( M - 1 \big)}{M}}    \\ \\  \approx        -   \frac{ \bigg[       1 + \frac{1}{N} + \mathcal{O} \big[ N^2 \big]     \bigg]  \bigg[ \big[ 1 +    \frac{1}{M} + \mathcal{O} \big[ M^2 \big]            \big]  - \chi_E - 1 \bigg]}{1 - \frac{1 + \epsilon \big( M - 1 \big)}{M}}  \\   \\ \approx        - \bigg[ 1 - \frac{1 + \epsilon \big( M - 1 \big)}{M} \bigg]^{-1}      \bigg[           1 + \frac{1}{N} + \mathcal{O} \big[ N^2 \big]                                  \bigg]      \bigg[         - \chi_E  +   \frac{1}{M} \\ + \mathcal{O} \big[ M^2 \big]                              \bigg]                      \\    \\ \approx       - \bigg[ 1 - \frac{1 + \epsilon \big( M - 1 \big)}{M} \bigg]^{-1}  \bigg[  - \chi_E +              \frac{1}{N} \frac{1}{M} + \mathcal{O} \big[ N^2   \big]  \mathcal{O} \big[ M^2   \big]   \\  - 2 \frac{\chi_E}{M}    + \bigg[ \frac{1}{\chi_E} \bigg]^2  +  \mathcal{O} \big[ N^2  \chi_E   \big]  \mathcal{O} \big[ M^2  \chi_E   \big]                                   \bigg]                    \\ \\ \equiv C_H \big[   N , M           \big] , 
\end{align*}

\noindent from which the inequality provided in the previous result above and the Helstrom inequality, implies,

\begin{align*}
    \mathrm{log} \bigg[ \frac{1}{N} \bigg] \big[ \mathrm{log} M - \chi_E - 1 \big] \overset{(\mathrm{Helstrom})}{\geq} C_H  \bigg[ 1 - \frac{1 + \epsilon \big( M - 1 \big)}{M}  \bigg]   , 
\end{align*}

\noindent from which we conclude the argument. \boxed{}

\bigskip

\noindent With all of the above results, we have demonstrated how the Fano's and Helstrom inequalities can be used to obtain the desired results, from which we conclude the argument. \boxed{}

\subsection{\textbf{Theorem} \textit{3}}

\noindent \textit{Proof of Theorem 3}. We reformulate the statement of the third main result in terms of the Holevo information, which is related to the classical-quantum polar codes and the symmetric information entropy. Fix the input distribution $P_X$ corresponding to each possible input $X$ that Alice provides to the Quantum channel. Namely, if there were to exist suitable protocols for Alice and Bob for messages containing infinitely many bits, given the classical input $X$ that Alice provides to the Quantum channel,

\begin{align*}
     \chi_B \big( p_X \big) > \chi_E \big( p_X \big)    , 
\end{align*}

\noindent corresponding to the fact that Eve's state, $E$, which she obtains by applying a CPTP map to Bob's input that is related to Alice's initial input $X$ is achievable at the transmission rate $R$ satisfying,

\begin{align*}
  0 \leq  R < \underset{P_X}{\mathrm{sup}}    \big\{     \chi_B \big( P_X \big) - \chi_E \big( P_X \big)         \big\}    ,
\end{align*}

\noindent as $n \longrightarrow + \infty$. Fix the mapping $W : \big\{ x \in X \big\} \mapsto \rho^{BE}_x$, where the image $\rho$ under $W$ can be factorized into states supported over Bob's and Eve's system, with,

\begin{align*}
 \rho^{BE}_X \equiv \underset{x \in X}{\prod} \rho^{BE}_x \equiv  \underset{x^{\prime} \in X \cap E }{\prod}  \bigg[ \bigg[ \underset{x \in X \cap B }{\prod} \rho^{B}_x \bigg]  \rho^E_{x^{\prime}} \bigg] \equiv \underset{x^{\prime} \in X \cap E }{\prod}  \bigg[ \bigg[ \underset{x \in X \cap B }{\prod} \rho^{B}_x \bigg] \mathcal{E} \big[  \rho^B_{x^{\prime}} \big]  \bigg]  . 
\end{align*}

\noindent To quantify the behavior for messages transmitted over the Quantum channel with infinitely many bits,

\begin{align*}
  \underset{n \longrightarrow + \infty}{\mathrm{lim}} R_n \longrightarrow R ,  
\end{align*}

\noindent write,

\begin{align*}
   \chi_B \big( P_X \big) =  S \bigg[ \underset{x \in X}{\sum} P_X \big( x \big) \rho^B_x  \bigg]    -     \underset{x \in X}{\sum} P_X \big( x \big) S \big( p^B_x \big)    , \\ \\     \chi_E \big( P_X \big) =    S \bigg[ \underset{x \in X}{\sum} P_X \big( x \big) \rho^E_x  \bigg]    -     \underset{x \in X}{\sum} P_X \big( x \big) S \big( p^E_x \big)     , 
\end{align*}

\noindent corresponding to the Holevo information of Bob and Eve, given an input distribution $p_X$ specified by Alice.

\bigskip

\noindent To argue that the set of possible $R$ can be achieved, denote,

\begin{align*}
    \mathcal{C} : = \underset{P_X}{\mathrm{sup}}    \big\{     \chi_B \big( P_X \big) - \chi_E \big( P_X \big)         \big\} , 
\end{align*}

\noindent from which, for $R \leq \mathcal{C}$:

\begin{itemize}
    \item[$\bullet$] \textit{Number of codewords mapped into the channel at rate $R$}. There are $2^{nR}$ messages that are mapped into channels with codewords that are of length $n$. Moreover, there exists a decoding for Bob, so that upon receiving a message from Eve his average decoding error is given by,

    \begin{align*}
      \textbf{P}_B \big[ \hat{M} \neq M \big] \leq \delta   ,  
    \end{align*}

    \noindent for $\delta$ taken to be sufficiently small.

    \bigskip

 \item[$\bullet$] \textit{Information leakage to Eve over the channel}. From the states in the image of $W$ for each $x$, the information leakage over the channel to Eve is given by the Holevo information,

 \begin{align*}
   \chi \big( M ; E^n \big) \equiv \textit{Holevo information between the measurement and Eve's channel over codewords} \\ \textit{ on the channel with length n}     .
 \end{align*}

\end{itemize}

\noindent As $n \longrightarrow + \infty$, each of the above quantities approach $0$.

\bigskip

\noindent We now make use of the list of properties provided in \textit{1.2}. That is, we argue that there exists protocols for Alice and Bob with the expressions obtained in the previous result for $p_{\mathrm{FA},\mathrm{E}}$. First, observe from previous arguments that there exists codes, from the Arikan polarization, that can be applied to $2^n$ uses of the approximated channel. Given cq-polar codes that have previously been characterized, a fraction approximately equal to the Holevo information $\chi_B \big( p_X \big)$ are good for Bob to accept. On the other hand, a fraction approximately equal to the Holevo information $\chi_E \big( p_X \big)$ are good for Eve to accept. As a result of the degradation assumption of the channel, hence approximately a fraction equal to $\chi_B \big( p_X \big) - \chi_E \big( p_X \big)$ of the synthesized channels are good for Bob, but bad for Eve, to accept. Straighforwardly, one can introduce a set of indices, $\mathcal{I}$, for which Bob can reliably decode the information being set by Alice.

Moreover, besides the construction of the cq-polar code one also can introduce a suitable encoder as follows. The encoder is characterized by the fact that, given the \textit{frozen} bits, namely the bits that are part of the transcript of Alice and Bob's communication over the public broadcast channel, revealing the frozen bits does not substantially leak information about the message. Quantitatively, the information leakage is given by the fact that Bob's error probability satisfies,

\begin{align*}
  P_{\mathrm{Error}, B} \leq \mathrm{O} \big( 2^{-N^{\beta}} \big)   ,
\end{align*}

\noindent for $\beta < 2^{-1}$ and $N$ taken sufficiently large. Therefore the reliability of the encoder approaches $0$. Given some $i \in \mathcal{I}$, the channel which Eve synthesizes has fidelity which is approximately $1$. Moreover, from arguments in the previous result making use of the Helstrom bound, per bit her success probability takes the form $2^{-1} + \mathrm{O} \big( 1 \big)$. This not only implies that her per bit leakage error is very small, but also that across the number of all such indices, namely $\big| \mathcal{I} \big| \approx N R$ bits, the supremum,

\begin{align*}
\underset{\textit{bits}}{\mathrm{sup}} \big\{ P_{\textit{Error}, bits, E} \big\} \equiv   P_{\textit{Error}, E}          \lesssim  \chi_B - \chi_E    \equiv   N \big( \chi_B - \chi_E \big)   , 
\end{align*}

\noindent of Eve's error probabilities is upper bounded by $\big( \chi_B - \chi_E \big)$ up to a constant. Hence the desired result follows from combining the previous observations, namely fixing some $\epsilon > 0$ and $N$ sufficiently large so that Bob's decoding error is at most $\epsilon$, in addition to the leakage of Eve's Holevo information being at most $\epsilon$. This implies that a suitable code can be constructed, at the previously defined rate $R$, so that reliability and secrecy both vanish as $\epsilon \longrightarrow 0$ and $N \longrightarrow + \infty$, from which we conclude the argument. \boxed{}

\subsection{\textbf{Corollary} \textit{1}}

\noindent \textit{Proof of Corollary 1}. The desired result follows from previous arguments provided by the author in {\color{blue}[48]}, specifically in \textbf{Corollary}, from which we conclude the argument. \boxed{}

\subsection{\textbf{Corollary} \textit{2}}

\noindent \textit{Proof of Corollary 2}. The desired form for Eve's constant, $C_{\mathrm{Eve}}$, straightforwardly follows from a consequence of arguments provided for the first main result, in which,

\[ \left\{\!\begin{array}{ll@{}>{{}}l}    C_{\mathrm{Eve}} \approx  1 - \frac{1 + \epsilon ( M - 1 )}{M}  \Longleftrightarrow M \text{ } \textit{small} , \\ \\      C_{\mathrm{Eve}} \backsimeq     \mathrm{log} \big[ \frac{1}{N} \big] \big[ \mathrm{log} M - \chi_E - 1 \big]  \Longleftrightarrow M \text{ } \textit{not small} ,              
\end{array}\right.
\] 

\noindent from which we conclude the argument. \boxed{}

\section{Conclusion}

\noindent In this work we demonstrated how to make use of Classical-Quantum codes that have previously been constructed in the literature, towards the purpose of fidelity pruning. In comparison to previous arguments due to the author that have been developed for pruning alphabets, the Tal et al., {\color{blue}[53]} approach allows one to not only guarantee some prescribed loss in the sum rate, but also an accompanying fidelity parameter. However, as one will recall in the current setting we compute the Holevo entropy, as well as Holevo sums, for demonstrating that $P_{\mathrm{FA},\mathrm{E}}$ satisfies a desired inequality. Through Classical-Quantum polar code constructions provided by Wilde and Guha, {\color{blue}[52]}, straightforwardly one can obtain the achievable rate that is equal to that of the sum-rate of the approximate channel. Additionally, in comparison to the Tal et al. approach, we are able to accommodate noncommuting density matrices by making use of the trace-distance, and fidelity, bounds, in comparison to total-variation distance bounds. In the future, it continues to remain of interest to explore topics relating to: Maurer-style public communication; cascaded CPTP channels; Eve's distinguishability and security under degradation; error exponents; or even information-theoretic security in hybrid Classical-Quantum networks.

\bigskip


\nocite{*}
\bibliography{sn-bibliography}

@article{Amr,
  author		= "Amr, A. and Villanueva, I.",
  title			= "Quantum one way vs. classical two way communication in XOR games",
  journal		= "Quantum Information Processing",
  volume		= "20",
  number		= "79",
  pages			= "",
  year			= "2021",
doi = ""
}

@article{Bannik,
  author		= "Bannik, T. and et al.",
  title			= "Bounding Quantum-Classical Separations for Classes of Nonlocal Games",
  journal		= "STACS",
  volume		= "12",
  number		= "",
  pages			= "1-12",
  year			= "2019",
doi = "https://doi.org/10.4230/LIPIcs.STACS.2019.12"
}

@article{Briet,
  author		= "Briet, J. and Buhrman, H. and Toner, B.",
  title			= "A generalized Grothendieck inequality and entanglement in XOR games",
  journal		= "Comm. Math. Phys.",
  volume		= "305",
  number		= "",
  pages			= "827-843",
  year			= "2011",
doi = "https://doi.org/ 10.1007/s00220-011-1280-3"
}

@article{Broadbent,
  author		= "Broadbent, A. and Methot, A.A.",
  title			= "On the power of non-local boxes",
  journal		= "Theoretical Computer Science",
  volume		= "358",
  number		= "",
  pages			= "3-14",
  year			= "2006",
doi = "https://doi.org/10.1016/j.tcs.2005.08.035"
}

@article{Brassard,
  author			= "Brassard, G. and Broadbent, A. and Tapp, A.",
  title		= "Quantum Pseudo-Telepathy",
journal = "Found. Phys.",
  volume		= "35",
  number		= "",
  pages			= "1877-1907",
  year			= "2005",
doi = "https://philpapers.org/rec/BRAQP"
}

@article{Benedetti,
  author		= "Benedetti, M. and Coyle, B. and Fiorentini, M. and Lubasch, M. and Rosenkranz, M.",
  title			= "{Variational Inference with a Quantum Computer}",
  journal		= "Phys Rev Applied",
  volume		= "16",
  number		= "044057",
  pages			= "",
  year			= "2021",
doi = "https://doi.org/10.1103/PhysRevApplied.16.044057"
}

@article{Bittel,
  author		= "Bittel, L. and Kliesch, M.",
  title			= "Training Variational Quantum Algorithms is NP-Hard. ",
  journal		= "Physical Review Letters",
  volume		= "127",
  number		= "120502",
  pages			= "",
  year			= "2021",
doi = "https://doi.org/10.1103/PhysRevLett.127.120502"
}

@article{Catani,
  author		= "Catani, L. and Faleiro, R. and Emeriau,P.E. and Mansfield,S. and Pappa, A.",
  title			= "Connecting XOR and XOR* games",
  journal		= "Phys. Rev. A",
  volume		= "109",
  number		= "012427",
  pages			= "",
  year			= "2024",
doi = "https://doi.org/10.1103/PhysRevA.109.012427"
}

@article{Chen,
  author		= "Chen, H. and Vives, M. and Metcalf, M.",
  title			= "Parametric amplification of an optomechanical quantum interconnect",
  journal		= "Physical Review Research",
  volume		= "4",
  number		= "043119",
  pages			= "",
  year			= "2022",
doi = "https://doi.org/10.1103/PhysRevResearch.4.043119"
}

@article{Cong,
  author		= "Cong, I. and Duan, L.",
  title			= " Quantum discriminant analysis for dimensionality reduction and classification",
  journal		= "New Journal of Physics",
  volume		= "18",
  number		= "073011",
  pages			= "",
  year			= "2016",
doi = "https://doi.org/10.1088/1367-2630/18/7/073011 "
}

@article{Cleve,
  author		= "Cleve, R. and Hoyer, P. and Toner, B. and Watrous, J.",
  title			= "Consequences and Limits of Nonlocal Strategies",
  journal		= "19th IEEE Annual Conference on Computational Complexity Proceedings",
  volume		= "",
  number		= "",
  pages			= "236-249 ",
  year			= "2004",
doi = "$\mathrm{https://doi.org/10.1109/CCC.2004.1313847}$"
}

@article{Culf,
  author		= "Culf, E. and Mousavi, H. and Spirig, T.",
  title			= "Approximation Algorithms for Noncommutative CSPs",
  journal		= "IEEE 65th Annual Symposium on Foundations of Computer Science (FOCS)",
  volume		= "",
  number		= "",
  pages			= "920-929",
  year			= "2024",
doi = "$\mathrm{https://doi.org/ 10.1109/FOCS61266.2024.00061}$"
}

@article{Cui,
  author		= "Cui, D. and  Malavolta, G. and Mehta, A. and Natarajan, A. and Paddock, C. and Schmidt, S. and Walter, M. and Zhang, T.",
  title			= "A Computational Tsireslson's Theorem for the Value of Compiled XOR games",
  journal		= "arXiv: 2402.17301",
  volume		= "",
  number		= "",
  pages			= "",
  year			= "2024",
doi = ""
}

@article{Doherty,
  author		= "Doherty, A.C. and Liang, Y.C. and Toner, B. and Wehner, S.",
  title			= "The Quantum Moment Problem and Bounds on Entangled Multi-Prover Games",
  journal		= "23rd Annual IEEE Conference on Computational Complexity",
  volume		= "8",
  number		= "",
  pages			= "",
  year			= "2018",
doi = ""
}

@article{Dmrota,
  author		= "Drmota, P. and Main, D. and Ainley, E.M. and Agrawal, A. and Araneda, G. and Nadlinger, Srinivas, R. and Cabello, A. and et al.",
  title			= "Experimental Quantum Advantage in the Odd-Cycle Game",
  journal		= "Phys. Rev. Lett.",
  volume		= "134",
  number		= "070201",
  pages			= "",
  year			= "2025",
doi = "https://doi.org/10.1103/PhysRevLett.134.070201"
}

@article{Ewe,
  author		= "Ewe, W-B. and Koh, D. E. and Goh, S. T. and Chu, H-S and Png, C. E.",
  title			= "Variational Quantum-Based Simulation of Waveguide Modes",
  journal		= "IEEE Transactions on Microwave Theory and Techniques",
  volume		= "70",
  number		= "5",
  pages			= "2517-2525",
  year			= "2022",
doi = "https://doi.org/10.1109/TMTT.2022.3151510"
}

@article{PE,
  author		= "Pierre-Emmanuel Emeriau, P-E. and Howard, M. and Mansfield, S.",
  title			= "Quantum Advantage in Information Retrieval",
  journal		= "PRX Quantum",
  volume		= "3",
  number		= "020307",
  pages			= "",
  year			= "2022",
doi = "\mathrm{ https://doi.org/10.1103/PRXQuantum.3.020307}"
}

@article{Faleiro,
  author		= "Faleiro, R.",
  title			= "Quantum strategies for simple 2-player XOR games",
  journal		= "Quantum Inf Process",
  volume		= "19",
  number		= "229",
  pages			= "",
  year			= "2020",
doi = "\mathrm{doi:10.1007/s11128-020-02717-2}"
}

@article{Garg,
  author		= "Garg, D. and Ikbal, S. and Srivastava, S.K. and Vishwakarma, H. and Karanam, H. and Subramaniam, L.V.",
  title			= "Quantum Embedding of Knowledge for Reasoning.",
  journal		= "Advance in Neural Information Processing Systems",
  volume		= "32",
  number		= "",
  pages			= "",
  year			= "2019",
doi = " "
}

@article{Genoni,
  author		= "Genoni, M.G. and Tufarelli, T.",
  title			= "Non-orthogonal bases for quantum metrology",
  journal		= "Journal of Physics A: Mathematical and Theoretical",
  volume		= "52",
  number		= "43",
  pages			= "",
  year			= "2019",
doi = "https://doi.org/10.1088/1751-8121/ab3fe0"
}

@article{Gidi,
  author		= "Gidi, J.A. and Candia, B. and Munoz-Moller, A.D. and Rojas, A. and Pereira, L. and Munoz, M. and Zambrano, L. and Delgado, A.",
  title			= "Stochastic optimization algorithms for quantum applications",
  journal		= "Phys.Rev.A",
  volume		= "108",
  number		= "032409",
  pages			= "",
  year			= "2023",
doi = "https://doi.org/10.1103/PhysRevA.108.032409"
}

@article{Givi,
  author		= "Givi, P. and Daley, A.J. and Mavriplis, D. and Malik, M.",
  title			= "Quantum Speedup for Aeroscience and Engineering",
  journal		= "AIAA",
  volume		= "58",
  number		= "8",
  pages			= "",
  year			= "2020",
doi = " "
}

@article{B,
  author		= "Helton, J.W. and Mousavi, H. and Nezhadi, S.S. and et al.",
  title			= "Synchronous Values of Games",
  journal		= "Ann. Henri Poincaré",
  volume		= "25",
  number		= "",
  pages			= "4357–4397",
  year			= "2024",
doi = "https://doi.org/10.1007/s00023-024-01426-1"
}

@article{Hadiashar,
  author		= "Hadiashar, S.B. and Nayak, A. and Sinha, P.",
  title			= "Optimal lower bounds for Quantum Learning via Information Theory",
  journal		= "IEEE Transactions on Information Theory",
  volume		= "70",
  number		= "3",
  pages			= "1876--1896",
  year			= "2024",
doi = "https://doi.org/10.1109/TIT.2023.3324527"
}

@article{Hur,
  author		= "Hur, T. and Kim, L. and Park, D.K.",
  title			= "Quantum convolutional neural network for classical data classification",
  journal		= "Quantum Machine Intelligence",
  volume		= "4",
  number		= "3",
  pages			= "",
  year			= "2022",
doi = "https://doi.org/10.1007/s42484-021-00061-x"
}

@article{Holmes,
  author		= "Holmes, Z. and Coble, N.J. and Sornborger, A.T. and Subasi, Y.",
  title			= "On nonlinear transformations in quantum computation",
  journal		= "Phys. Rev. Research",
  volume		= "5",
  number		= "013105",
  pages			= "",
  year			= "2023",
doi = "https://doi.org/10.1103/PhysRevResearch.5.013105"
}

@article{Jing,
  author		= "Jing, H. and Wang, Y. and Li, Y.",
  title			= "Data-Driven Quantum Approximate Optimization Algorithm for Cyber-Physical Power Systems",
  journal		= "arXiv: 2204.00738",
  volume		= "",
  number		= "",
  pages			= "",
  year			= "2022",
doi = "https://doi.org/10.48550/arXiv.2204.00738"
}

@article{Junge,
  author		= "Junge, M. and Palazuelos, C.",
  title			= " On the power of quantum entanglement in multipartite quantum XOR games",
  journal		= "Journal of the London Mathematical Society",
  volume		= "110",
  number		= "5",
  pages			= "",
  year			= "2024",
doi = ""
}

@article{Kubo,
  author		= "Kubo, K. and Nakagawa, Y.O. and Endo, S. and Nagayama, S.",
  title			= "Variational quantum simulations of stochastic differential equations",
  journal		= "Physical Review A",
  volume		= "103",
  number		= "052425",
  pages			= "",
  year			= "2021",
doi = "https://doi.org/10.1103/PhysRevA.103.052425 "
}

@article{Kribs,
  author		= "Kribs, D.W.",
  title			= "A quantum computing primer for operator theorists",
  journal		= "Linear Algebra and its Applications",
  volume		= "400",
  number		= "",
  pages			= "147-167",
  year			= "2005",
doi = "https://doi.org/10.48550/arXiv.math/0404553 "
}

@article{Li,
  author		= "Li, R. Y. and Di Felice, R. and Rohs, R. and Lidar, D.A.",
  title			= "Quantum annealing versus classical machine learning applied to a simplied computational biology problem",
  journal		= "npj Quantum Information",
  volume		= "4",
  number		= "14",
  pages			= "",
  year			= "2008",
doi = " https://doi.org/10.1038/s41534-018-0060-8"
}

@article{Mahdian,
  author		= "Mahdian, M. and Yeganeh, H.D.",
  title			= "Toward a quantum computing algorithm to quantify classical and quantum correlation of system states",
  journal		= "Quantum Information Processing",
  volume		= "20",
  number		= "393",
  pages			= "",
  year			= "2021",
doi = " https://doi.org/10.1007/s11128-021-03331-6
 "
}

@article{Maldonado,
  author		= "Maldonado, T.J. and Flick, J. and Krastanov, S. and Galda, A.",
  title			= "Error rate reduction of single-qubit gates via noise-aware decomposition into native gates",
  journal		= "Scientific Reports",
  volume		= "12",
  number		= "6379",
  pages			= "",
  year			= "2022",
doi = "https://doi.org/10.1038/s41598-022-10339-0 "
}

@article{Manby,
  author		= "Manby, F.R. and Stella, M. and Goodpaster, J.D. and Miller, T.F.",
  title			= "A Simple, Exact Density-Functional-Theory Embedding Scheme",
  journal		= "Journal of Chemical Theory and Computation",
  volume		= "8",
  number		= "8",
  pages			= "2564-2568",
  year			= "2012",
doi = " https://doi.org/10.1021/ct300544e"
}

@article{Maurer,
  author		= "Maurer, U.",
  title			= "Perfect Cryptographic Security from Partially Independent Channels",
  journal		= "Proc. 23rd ACM Symposium on Theory of Computing — STOC",
  volume		= "",
  number		= " ",
  pages			= "561–572",
  year			= "1991",
doi = "https://crypto.ethz.ch/publications/Maurer91b.html"
}

@article{Mensa,
  author		= "Mensa, S. and Sahin, E. and Tacchino, F. and Barkoutsos, P.K. and Tavernelli, I.",
  title			= "Quantum Machine Learning Framework for Virtual Screening in Drug Discovery: a Prospective Quantum Advantage",
  journal		= "Mach. Learn.: Sci. Technol.",
  volume		= "4",
  number		= "015023",
  pages			= "",
  year			= "2023",
doi = " https://doi.org/10.1088/2632-2153/acb900"
}

@article{Sheng,
  author		= "Nan Sheng, H.M. and Govono, M. and Galli, G.",
  title			= "Quantum Embedding Theory for Strongly-Correlated States in Materials",
  journal		= "J. Chem. Theory Comput.",
  volume		= "17",
  number		= "4",
  pages			= "2116-2125",
  year			= "2021",
doi = " https://doi.org/10.1021/acs.jctc.0c01258"
}

@article{Ostrev,
  author		= "Ostrev, D.",
  title			= "The structure of nearly-optimal quantum strategies for the non-local XOR games ",
  journal		= "Quantum Information and Computation",
  volume		= "16",
  number		= "13-14",
  pages			= "1191-1211",
  year			= "2016",
doi = "https://doi.org/10.26421/QIC16.13-14-6"
}

@article{Ostrev2,
  author		= "Ostrev, D.",
  title			= "Composable, Unconditionally Secure Message Authentication without any Secret Key",
  journal		= "IEEE International Symposium on Information Theory",
  volume		= "10",
  number		= "1109",
  pages			= "622-626",
  year			= "2019",
doi = "https://doi.org/10.1109/ISIT.2019.8849510"
}

@article{Paine,
  author		= "Paine, A.E. and Elfving, V.E. and Kyriienko, O.",
  title			= "Quantum Kernel Methods for Solving Differential Equations",
  journal		= "Physical Review A",
  volume		= "107",
  number		= "032428",
  pages			= "",
  year			= "2023",
doi = "https://doi.org/10.1103/PhysRevA.107.032428"
}

@article{Paudel,
  author		= "Paudel, H.P. and Syamlal, M. and Crawford, S.E. and Lee, Y-L and Shugayev, R.A. and Lu, P. and Ohodnicki, P.R. and Mollot, D. and Duan, Y.",
  title			= "Quantum Computing and Simulations for Energy Applications: Review and Perspective",
  journal		= "ACS Eng. Au",
  volume		= "3",
  number		= "",
  pages			= "151-196",
  year			= "2022",
doi = "https://doi.org/10.1021/acsengineeringau.1c00033 
"
}

@article{Przhiyalkovskiy,
  author		= "Przhiyalkovskiy, Y.V.",
  title			= "Quantum process in probability representation of quantum mechanics",
  journal		= "Journal of Physics A: Mathematical and Theoretical",
  volume		= "55",
  number		= "085301",
  pages			= "",
  year			= "2022",
doi = " https://doi.org/10.1088/1751-8121/ac4b15 "
}

@article{Perc,
  author		= "Perc, M.",
  title			= "Statistical physics of human cooperation",
  journal		= "Physics Reports",
  volume		= "687",
  number		= "",
  pages			= "1-51",
  year			= "2017",
doi = "https://papers.ssrn.com/sol3/papers.cfm?abstract_id=2972841"
}

@article{Ravi,
  author		= "Ravishankar Ramanathan, R. and Augusiak, R. and Murta, G.",
  title			= "Generalized XOR games with $d$ outcomes and the task of nonlocal computation",
  journal		= "Phys. Rev. A",
  volume		= "93",
  number		= "022333",
  pages			= "",
  year			= "2016",
doi = "https://doi.org/10.1103/PhysRevA.93.022333"
}

@article{RigasP,
  author		= "Rigas, P.",
  title			= "Optimal, and approximately optimal, quantum strategies for $\mathrm{XOR^{*}}$ and $\mathrm{FFL}$ games",
  journal		= "arXiv: 2311.12887 (submitted)",
  volume		= "",
  number		= "",
  pages			= "",
  year			= "2023",
doi = ""
}

@article{Rigas,
  author		= "Rigas, P.",
  title			= "Variational quantum algorithm for measurement extraction from the Navier-Stokes, Einstein, Maxwell, B-type, Lin-Tsien, Camassa-Holm, DSW, H-S, KdV-B, non-homogeneous KdV, generalized KdV, KdV, translational KdV, sKdV, B-L and Airy equations",
  journal		= "arXiv: 2209.07714 (submitted)",
  volume		= "",
  number		= "",
  pages			= "",
  year			= "2025",
doi = "
https://doi.org/10.48550/arXiv.2209.07714"
}

@article{Rigas22,
  author		= "Rigas, P.",
  title			= "Quantum error bounds, optimality, and duality gaps for multiplayer XOR, XOR*, compiled XOR, XOR*, and strong parallel repetition of XOR, XOR*, and FFL games",
  journal		= "arXiv: 2209.07714 (submitted)",
  volume		= "",
  number		= "",
  pages			= "",
  year			= "2025",
doi = "
https://doi.org/10.48550/arXiv.2209.07714"
}

@article{Rigas23,
  author		= "Rigas, P.",
  title			= "
Error correction, authentication, and false acceptance, probabilities for communication over noisy quantum channels: converse upper bounds on the bit transmission rate",
  journal		= "arXiv: 2507.03035 (submitted)",
  volume		= "",
  number		= "",
  pages			= "",
  year			= "2025",
doi = "
https://doi.org/10.48550/arXiv.2507.03035

"
}

@article{Rigas24,
  author		= "Rigas, P.",
  title			= "Parallel repetition of expanded, and multiplayer, Quantum games: anchoring, optimal values, generalized error bounds, dependency-breaking as symmetry-breaking",
  journal		= "arXiv: 2508.09380 (submitted)",
  volume		= "",
  number		= "",
  pages			= "",
  year			= "2025",
doi = "
https://doi.org/10.48550/arXiv.2508.09380
"
}

@article{Rosicka,
  author		= "Roscika, M. and Mazurek, P. and Grudka, A. and Horodecki, M. ",
  title			= "Generalized XOR non-locality games with graph description on a square lattice",
  journal		= "Journal of Phys A: Math. Theor.",
  volume		= "53",
  number		= "265302",
  pages			= "",
  year			= "2020",
doi = "https://doi.org/10.1088/1751-8121/ab8f3e"
}

@article{Slofstra,
  author		= "Slofstra, W.",
  title			= "Lower bounds on the entanglement needed to play XOR non-local games",
  journal		= "Journal of Mathematical Physics",
  volume		= "52",
  number		= "10",
  pages			= "102202",
  year			= "2011",
doi = " https://doi.org/10.1063/1.3652924"
}

@article{Wilde,
  author		= "Wilde, M. and Guha, S.",
  title			= "Polar codes for classical-quantum channels",
  journal		= "IEEE Transactions on
Information Theory",
  volume		= "59",
  number		= "2",
  pages			= "1175-1187",
  year			= "2013",
doi = "https://doi.org/10.1109/TIT.2012.2218792 "
}

@article{Sharov,
  author		= "Tal, I. and Sharov, A. and Vardy, A.",
  title			= "Constructing polar codes for non-binary alphabets and MACs",
  journal		= "IEEE International Symposium on Information Theory Proceedings, Cambridge, MA, USA",
  volume		= "",
  number		= "",
  pages			= "2132-2136",
  year			= "2012",
doi = "https://doi.org/10.1109/ISIT.2012.6283739 "
}

\end{document}